\documentclass[iop]{emulateapj}

\usepackage{amsmath}
\usepackage{hyperref}

\begin{document}


\title{On the role of the Kerr-Newman black hole in the GeV emission of long gamma-ray bursts}

\author{R.~Ruffini\altaffilmark{1,2,3,4},
R.~Moradi\altaffilmark{1,3},
J.~A.~Rueda\altaffilmark{1,3,4}
Y.~Wang\altaffilmark{1,3},
Y.~Aimuratov\altaffilmark{1,2,5},
L.~Becerra\altaffilmark{1,3},
C.~L.~Bianco\altaffilmark{1,3},
Y.-C.~Chen\altaffilmark{1,3},
C.~Cherubini \altaffilmark{3,6,7},
S.~Filippi \altaffilmark{3,6,7},
M.~Karlica \altaffilmark{1,3}
G.~J.~Mathews\altaffilmark{8},
M.~Muccino\altaffilmark{7},
G.~B.~Pisani\altaffilmark{1,3},
D.~Primorac\altaffilmark{1,3},
S.~S.~Xue \altaffilmark{1,3}
}

\altaffiltext{1}{ICRA and Dipartimento di Fisica, Universit\`a  di Roma ``La Sapienza'', Piazzale Aldo Moro 5, I-00185 Roma, Italy}
\altaffiltext{2}{Universit\'e de Nice Sophia-Antipolis, Grand Ch\^ateau Parc Valrose, Nice, CEDEX 2, France} 
\altaffiltext{3}{International Center for Relativistic Astrophysics Network, Piazza della Repubblica 10, I-65122 Pescara, Italy}
\altaffiltext{4}{ICRANet-Rio, Centro Brasileiro de Pesquisas F\'isicas, Rua Dr. Xavier Sigaud 150, 22290--180 Rio de Janeiro, Brazil}
\altaffiltext{5}{Fesenkov Astrophysical Institute, Observatory 23, 050020 Almaty, Kazakhstan}
\altaffiltext{6}{Department of Engineering, University Campus Bio-Medico of Rome, Nonlinear Physics and Mathematical Modeling Lab, Via Alvaro del Portillo 21, 00128 Rome, Italy}
\altaffiltext{7}{International Center for Relativistic Astrophysics–ICRA, University Campus Bio-Medico of Rome, Via Alvaro del Portillo 21, I-00128 Rome, Italy}
\altaffiltext{8}{Center for Astrophysics, Department of Physics, University of Notre Dame, Notre Dame, IN, 46556, USA}


\begin{abstract}
{X-ray Flashes (XRFs), binary-driven hypernovae (BdHNe) are long GRB subclasses with progenitor a CO$_{\rm core}$, undergoing a supernova (SN) explosion and  hypercritically accreting in a tight binary system onto a companion neutron star (NS) or black hole (BH). In XRFs the NS does not reach by accretion the critical mass and no BH is formed. In BdHNe I, with shorter binary periods, the NS gravitationally collapses and leads to a new born BH. In BdHNe II the accretion on an already formed BH leads to a more massive BH. We assume that the GeV emission observed by \textit{Fermi}-LAT originates from the rotational energy of the BH. Consequently, we  verify that, as expected, in XRFs no GeV emission is observed. In $16$ BdHNe I and $5$ BdHNe II, within the boresight angle of LAT, the integrated GeV emission allows to estimate the initial mass and spin of the BH. In the remaining $27$ sources in the plane of the binary system no GeV emission occurs, hampered by the presence of the HN ejecta. From the ratio, $21/48$, we infer a new asymmetric morphology for the BdHNe reminiscent of the one observed in active galactic nuclei (AGN): the GeV emission occurs within a cone of half-opening angle $\approx 60^{\circ}$ from the normal to the orbital plane of the binary progenitor. The transparency condition requires a Lorentz factor $\Gamma \sim 1500$ on the source of GeV emission. The GeV luminosity in the rest-frame of the source follows a universal power-law with index of $-1.20 \pm 0.04$, allowing to estimate the spin-down rate of the BH}.
\end{abstract} 

\keywords{gamma-ray bursts: general --- binaries: general --- stars: neutron --- supernovae: general --- black hole physics --- hydrodynamics}

\section{Introduction}\label{sec:1}

{The earliest evidence for high energy radiation over 100 MeV from GRBs where the detection by the Energetic Gamma-Ray Experiment Telescope (\textit{EGRET)},  operating in the energy range $\sim$ 20 MeV-30 GeV, onboard of the Compton Gamma-Ray Observatory (\textit{CGRO}, 1991-2000) where the detection was triggered by the Burst And Transient Source Experiment (\textit{BATSE}), operating in energy range of $\sim$ 20-2000 keV.}

{\textit{EGRET} has detected five GRBs, from our understanding today, all 5 GRBs where long GRBs: GRB 910503, GRB 910601, GRB 930131, GRB 940217, and GRB 940301 \citep[see e.g.][and references therein]{1996MmSAI..67..161K}. Unfortunately no redshift where known at the time.}

{A new era started with the launch of \textit{AGILE} in 2007 \citep{2009A&A...502..995T} with onboard Gamma Ray Imaging Detector (\textit{GRID}) operating in the 30~MeV-50~GeV energy range and the launch in June 2008 of the \textit{Fermi} satellite, having onboard the Large Area Telescope (LAT) operating from early August 2008 \citep{2009ApJ...697.1071A}.}

{\textit{AGILE-GRID} detected the first long GRB with emission above 100 MeV with photometric redshift, $z=1.8$, GRB 080514B, \citep{2008A&A...491L..25G} this event was followed four months later by the detection of GRB 080916C \citep{2009A&A...498...89G}  by \textit{Fermi}  with one of the largest isotropic energy ever detected, $E_{iso}= (4.07 \pm 0.86) \times 10^{54}$~erg and photometric redshift $z= 4.35$. These were followed by a multitude number of long GRBs by LAT with both GeV emission and with well-defined cosmological redshift, $z$.} 

{One year later the first observation of a short GRB  was done by \textit{AGILE}, GRB 090510A, with spectroscopic redshift $z=0.903$ and $E_{iso}= (3.95 \pm 0.21) \times 10^{52}$~erg and $E_{LAT}= (5.78 \pm 0.60) \times 10^{52}$~erg. On the ground of the observed energetics of this source, and its spectral properties, we proposed that in this short GRB we witness the birth of the Kerr-Newman black hole (BH), being the GeV emission the signature of this event \citep{2016ApJ...831..178R}.}

{In this article we address the role of the Kerr-Newman BH in explaining the GeV emission of all long GRBs and, as a by product, we  determine the initial spin and the mass of the BH. In order to do this it is necessary to approach this problem within a new model characterized by the process occurring in the binary-driven hypernova (BdHN) model \citep[see e.g.][and references therein]{2018ApJ...852...53R}. This will allow to answer some of the unsolved issues posed by the traditional GRB model \citep[see, e.g.][]{2018arXiv180401524N}.}

The traditional model {assumes that} all long-GRBs occur {from a single} ultra-relativistic jetted emission originating from a BH. {There is a single ultra relativistic blast wave \citep{1976PhFl...19.1130B} extending from the prompt to the latest emission of the afterglow and including, when present, the GeV emission.} {This traditional model has been well illustrated in a series of reviews \citep[see e.g.][]{1993ApJ...405..273W,1998ApJ...494L..45P,1999ApJ...524..262M,1999PhR...314..575P,Meszaros:2001vr,2004RvMP...76.1143P,2009MNRAS.396.1163Z,2009MNRAS.400L..75K,2013ApJ...764..179B, 2015MNRAS.450.1077B,2013RSPTA.37120273P}. This model is by now in contrast with observational facts also recognized by some of the original proposers of the model \citep[see e.g. ][]{2015MNRAS.454.1073B}}.
 
{Our alternative model differs significantly:}
{
I) A first difference is that long GRBs have been recognized  not to have a single common origin from a BH as assumed in the traditional approach. We have classified all GRBs in different subclasses on the basis of their energetics, spectra and duration, $T_{90}$, all expressed in the rest-frame of the source. Only in some of these subclasses the presence of a BH occurs \citep[see e.g.][]{2016ApJ...832..136R,2017arXiv171205001R}.
}

{
II) A second difference refers to the traditional assumption of a single ultra-relativistic process extending from the prompt phase all the way to the GeV emission. This contrasts with clear model-independent observational constraints. The plateau phase following the prompt radiation, presents a variety of processes, including hard and soft X-ray flares. Indeed, a model-independent approach constrains their Lorentz factor to $\Gamma \lesssim 2$ \citep{2018ApJ...852...53R}. The ultra-relativistic prompt radiation phase lasts less than $20$~s \citep[see e.g.][]{2017arXiv171205001R}. Our approach differs from the traditional one that sees the GeV radiation extending from ultra-relativistic phase into the plateau and afterglow \citep{2009MNRAS.400L..75K, 2009A&A...498..677B, Razzaque:2009xza}. This traditional approach is in contrast with the above limit on the Lorentz factor of the plateau. It has been also shown, moreover, that the mildly-relativistic phase of the plateau extends all the way to the late phase of afterglow \citep{2017arXiv171205000R}. For this reason in our alternative approach we focus on an ultra-relativistic proton-electron outflow originating from the BH and not at all related either to the plateau nor to the afterglow and conceptually different from the prompt phase. In this first article we indicate how to relate the GeV radiation to the presence of the BH. We then verify that the spin and mass of the BH, so determined, are sufficient to explain the observed energetics of the GeV emission.
}

{
III) A third difference is that the traditional model addresses a GRB based essentially on a single object neglecting the role of binary systems which we show characterize different GRB subclasses and allow as well with their dynamics to follow, in some of them, the birth of the BH. In particular we show in this article how the presence of the hypercritical accretion of the supernova (SN) originating from a CO$_{\rm core}$ on a companion neutron star (NS) or BH are necessary to explain basic properties of long GRBs, including the presence or absence of the GeV emission.
}

{
The main goal of our approach is first to estimate the integrated energy of the GeV emission for each GRB, $E_{\rm LAT}$. We assume that this energy originates from the newly formed Kerr-Newman BH in the BdHN and determine consequently the mass and spin of the BH and infer the astrophysical implication.
}

{
It has been well recognized in recent years that the GeV flux expressed as a function of the arrival time follows precise power-laws with index $\alpha=1.2 \pm 0.2$ \citep{2009MNRAS.400L..75K}, $\alpha=1.2 \pm 0.4$ \citep{2017ApJ...837...13P}. In our approach we adopt an alternative interpretation of these power-laws; instead of using the flux expressed in arrival time we use the luminosity expressed in the rest-frame of the source. This allows us to relate the GeV radiation to the slowing down of the BH spin.
}

{From the theoretical point of view we rely on three main topics of our previous research:}

{
a) The mass-energy formula of the Kerr BH \citet{1970PhRvL..25.1596C}, the one of the Kerr-Newman BH, \citet{1971PhRvD...4.3552C} and \citet{1971PhRvL..26.1344H, Hawking:1971vc} (in $c=G=1$ units), 
\begin{subequations}
\renewcommand{\theequation}{\theparentequation.\arabic{equation}}
\begin{align}
\label{aone}
M^2 = \frac{J^2}{4 M^2_{irr}}+\left(\frac{Q^2}{4 M_{irr}}+M_{irr}\right)^2,\\
S = 16 \pi M^2_{irr}    \\ 
\delta S = 32 \pi  M_{irr} \delta M_{irr}  \geq 0 ,
\end{align}
\end{subequations}
where $Q$, $J$, $M$, $M_{irr}$ and $S$ are the charge, angular momentum, mass, irreducible mass and horizon surface area of the BH. We  assume in this article that the contribution of the electromagnetic energy to the mass of the BH can be neglected. We will return in a forthcoming article to the essential role of the electromagnetic energy to explain the high-energy emission including the GeV. 
}

{
b) We  also rely on the description of hypercritical accretion recently summarized in three extended articles \citep[see e.g.][]{2016ApJ...832..136R,2016ApJ...833..107B,2018ApJ...852...53R} and the fundamental role of neutrinos \citep[see e.g.][]{2018ApJ...852..120B}.
}
\begin{figure*}
    \centering
\includegraphics[width=0.49\hsize,clip]{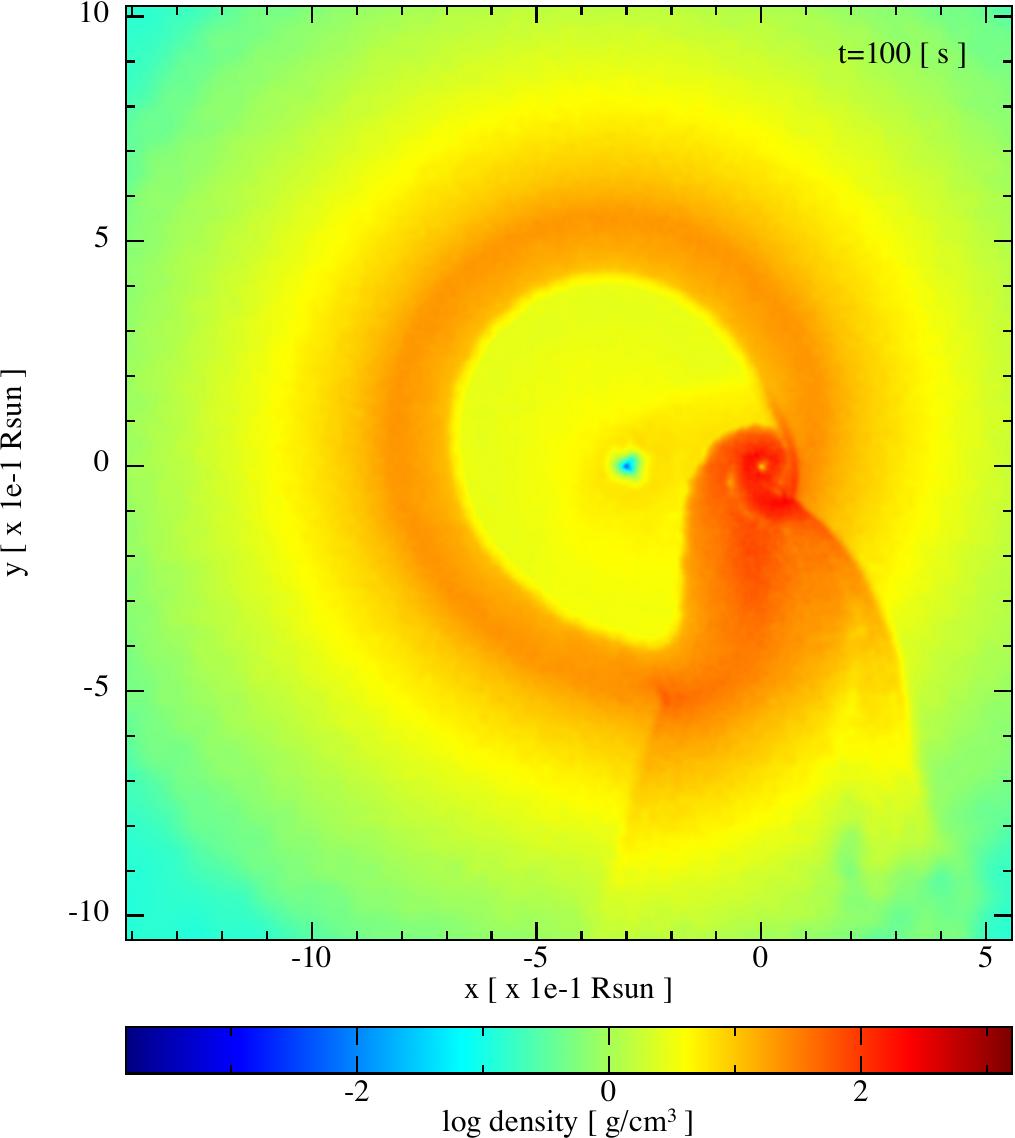}\includegraphics[width=0.49\hsize,clip]{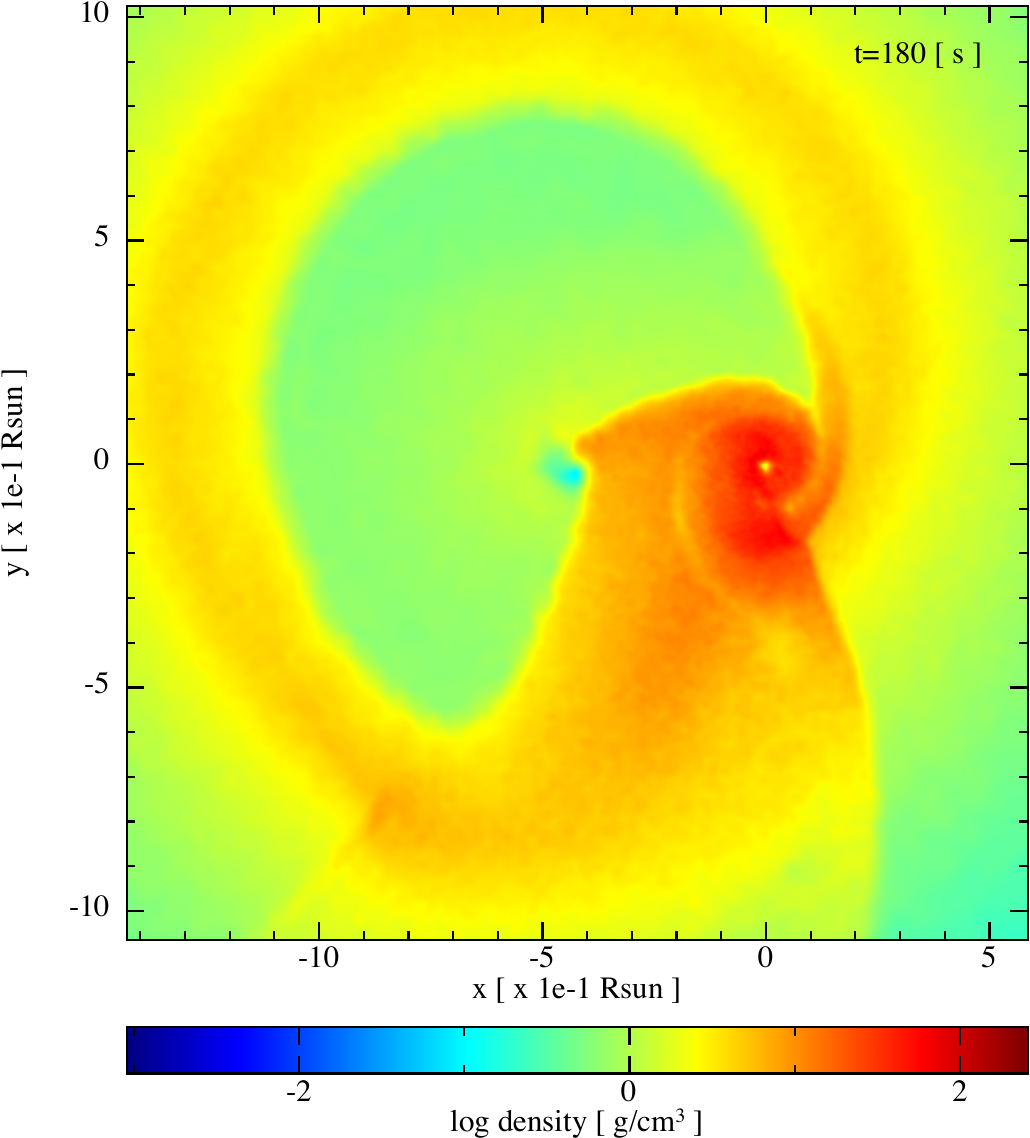}
\includegraphics[width=0.49\hsize,clip]{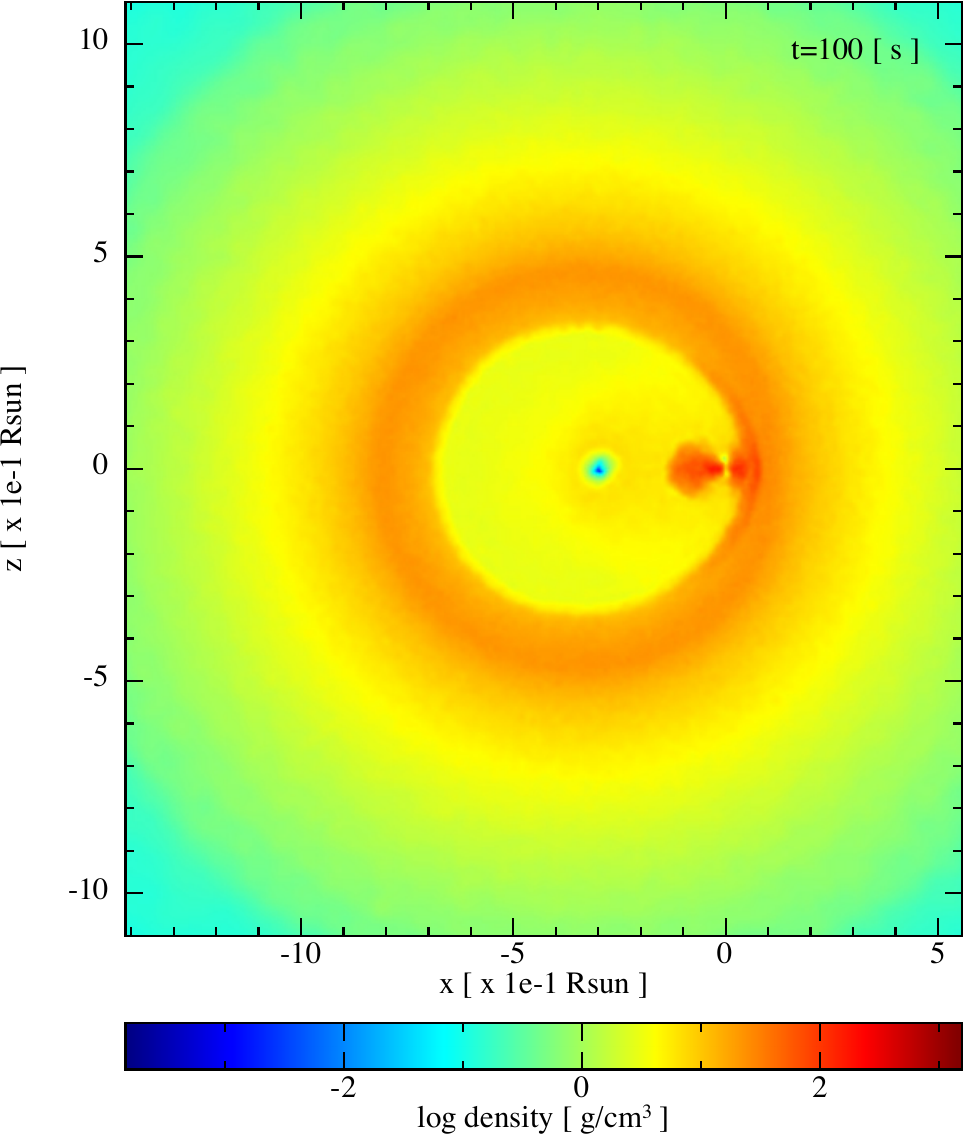}\includegraphics[width=0.49\hsize,clip]{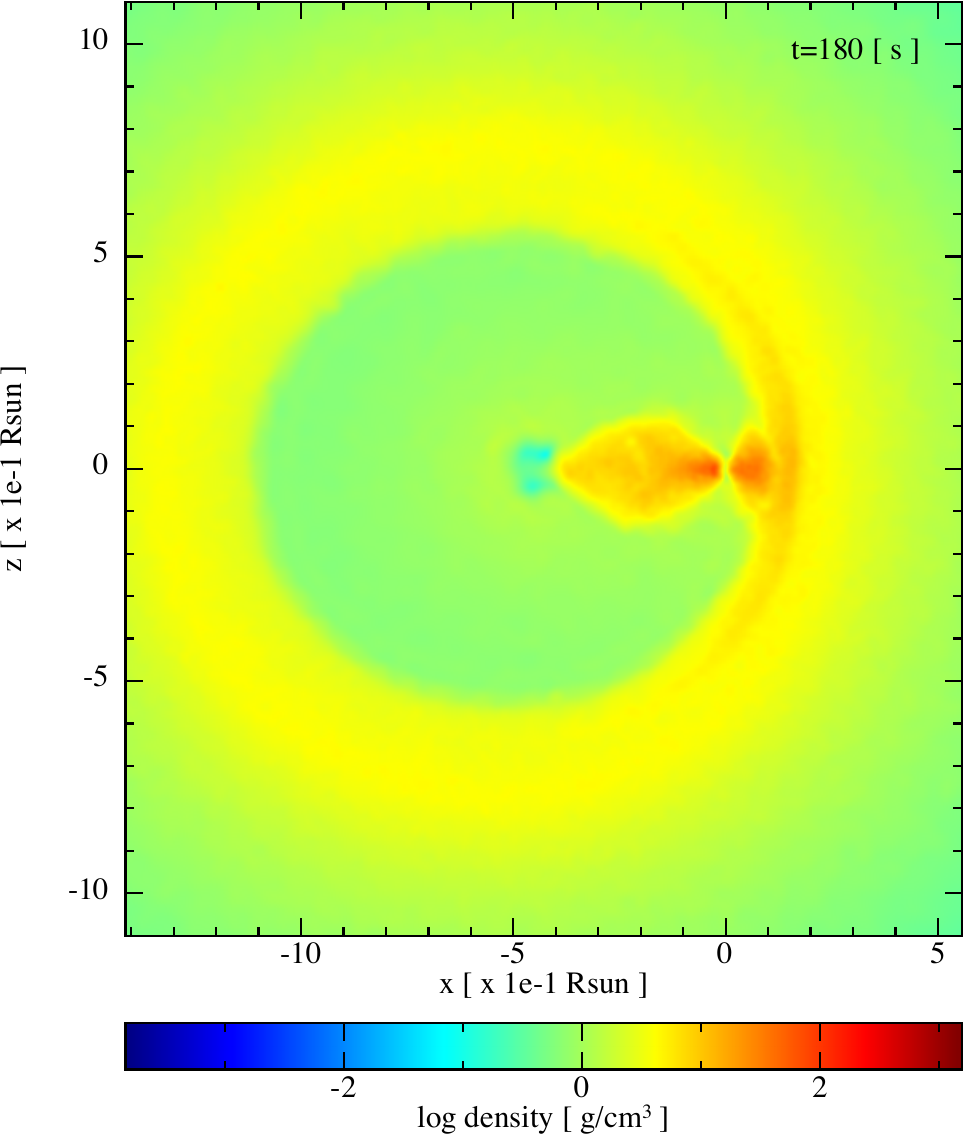}
\caption{{A selected SPH simulation from \citet{2018arXiv180304356B} of the exploding CO$_{\rm core}$ as SN in presence of a companion NS: Model ``25M1p1e'' (see table~2 there). The CO$_{\rm core}$ is taken from the $25~M_\odot$ zero-age main-sequence (ZAMS) progenitor which leads to a pre-SN CO$_{\rm core}$ mass $M_{\rm CO}=6.85~M_\odot$. The mass of the NS companion is $M_{\rm NS}=2~M_\odot$. The initial orbital period is of approximately $5$~min. The upper panel shows the mass density on the binary equatorial plane at two selected times from the SN shock breakout ($t=0$ of the simulation), $100$~s and $180$~s. The lower panel corresponds to the plane orthogonal to the binary equatorial plane. The reference system has been rotated and translated such that the x-axis is along the line that joins the binary stars and the origin is at the NS position. It can be seen that the SN ejecta particles circularize around the NS and form an accretion disk.}}\label{fig:SPHsimulation}
\end{figure*}

{
c) We finally rely on the three-dimensional (3D) simulations performed with a new SPH code needed to reconstruct of the morphology of the BdHNe as a function of the viewing angle \citep[see e.g.][and also Fig.~\ref{fig:SPHsimulation}]{2018arXiv180304356B}.
}

The aim of this article is to use the GeV emission of \textit{Fermi}-LAT for further differentiating the eight different families of GRBs presented in \citet{2016ApJ...832..136R}, in particular the difference between XRFs and BdHNe.

We first give in section~\ref{sec:2} an outline of the eight GRB subclasses presented in \citet{2016ApJ...832..136R} with a brief summary of their initial states (\textit{in-state}), their out-state, their energetics and spectral properties both in the gamma-rays and in the GeV emission.

In section~\ref{sec:3} we give a summary of observational properties of $48$ XRFs observed after the launch of \textit{Fermi}, including their cosmological redshifts, $E_{\rm p,i}$ of their spectrum, their isotropic energy $E_{\rm iso}$, the Fermi GCN and the boresight angle of \textit{Fermi}-LAT (if observed). As predicted, in none of them GeV emission is observed. 

In section~\ref{sec:4} we analyze the properties of the GeV emission in BdHNe. We consider only the ones which have the boresight angle of \textit{Fermi}-LAT less than $75^{\circ}$ at the time of the trigger. We give the details of the $21$ BdHNe {with observed GeV radiation}, out of the $329$ BdHNe with known cosmological redshift \citep{2016ApJ...833..159P}. For each, the cosmological redshift, the $E_{\rm p,i}$ of the spectrum, the $E_{\rm iso}$ of the source, the \textit{Fermi} GCN, the boresight angle, the $E_{\rm LAT}$, the likelihood test statistic (TS) and some distinguishing properties are given in Table~\ref{tab:cb}. 

In section~\ref{sec:5} we {determine} the values of the mass and spin of the BH adopting two {limiting} NS nuclear relativistic equations of state (EOS). {This leads to identify the subfamily of $16$ BdHNe of type I (BdHNe I), see Table~\ref{tab:dd} in which the process of hypercritical accretion of the SN occurs on a companion NS which gives origin to the newly formed BH. Most interesting is the fact that this leads to a BH with a mass very close to the NS critical mass and with spin systematically smaller that $0.71$}. 

In section~\ref{sec:6} we address the {subfamily of BdHNe of type II (BdHNe II)}, which originates in a binary progenitor composed of a CO$_{\rm core}$ and {a BH companion binary}, we give $5$ examples in Table~\ref{tab:2b}, of these systems which are  the most energetic observed BdHNe. {This brings to the total of $21$ BdHNe with observed GeV emission.} 

In section~\ref{sec:10} the cosmological redshift, the $E_{\rm p,i}$ of the spectrum, the $E_{\rm iso}$ of the source, the Fermi GCN, the boresight angle and some distinguishing properties of $27$ BdHNe  without GeV emission are given. We explain the nature of the these BdHNe in terms of a novel morphology of the binary system. The unexpected result appeared that the geometry of BdHNe, in some respect, is similar to the one of active galactic nuclei (AGN). In the case of BdHNe the GeV emission is constrained by the HN ejecta to be observable only through conical region normal to the orbital binary plane. When the BdHNe are seen along the orbital plane of the binary progenitor then the GeV emission is scattered by the HN ejecta and only the Gamma-ray flare, the X-ray flare and the X-ray plateau remain observable. {From the ratio of $21/48$ we conclude:} When the GRB is observed in a conical region of approximately $60^{\circ}$ around the normal to the plane of the binary progenitors, then all the emissions are observable, namely the X-ray, the gamma-ray and also the GeV emission. {For larger inclination angle the GeV radiation is absorbed by the HN ejecta and flaring activities are observed following the prompt radiation.}
 
In section~\ref{sec:11} we visualize this novel geometry using the result of recent smoothed-particle-hydrodynamics (SPH) numerical simulations at the moment of BH formation in a BdHN \citep{2018arXiv180304356B}.

{
In section~\ref{sec:7}, we compute the Lorentz Gamma factor of the GeV emission to fulfill the condition of  transparency, the obtained value is in the order of $\Gamma \sim 1500$. This is an additional  prove that GeV emission occurs in  an ultra-relativistic regime which differs from the mildly relativistic regime originating the plateau and afteglow.
}

{
In section~\ref{sec:8} for each of the $21$ BdHNe  we provide the $0.1$--$100$~GeV luminosity light-curves as a function of the time in the rest-frame of the source. {We obtain a power-law fit $L_n=A_n t^{-1.20\pm0.04}$~erg~s$^{-1}$ and report} the value of amplitude $A_n$ and the luminosity at $10$~s from the beginning of the prompt radiation, $L_{\rm 10s}$, with their associated uncertainties. We also provide a correlation between $L_{\rm 10s}$ and $E_{\rm iso}$.
}

{
In section~\ref{sec:9} we use the two aforementioned results, namely 1) that the GeV energetic is explained by the rotational energy of the BH and 2) the power-law evolution of the $0.1$--$100$~GeV luminosity, to infer the slowing down rate of the BH spin .
}

We finally proceed to the general conclusions in section~\ref{sec:12}.

Before proceeding we indicate in Table~\ref{acronyms} the alphabetic ordered list of acronyms used in this work.
\begin{table}
\centering
\begin{tabular}{lc}
\hline\hline
Extended wording & Acronym \\
\hline
Binary-driven hypernova & BdHN \\
Black hole                    & BH \\
Carbon-oxygen core      & CO$_{\rm core}$ \\ 
Gamma-ray burst         & GRB \\
Gamma-ray flash          & GRF \\
gamma-ray flash kilonovae & GR-K \\
Massive neutron star     & M-NS \\
Neutron star                & NS \\
New neutron star          & $\nu$NS \\
Short gamma-ray burst  & S-GRB \\
Short gamma-ray flash  & S-GRF \\
Supernova                  & SN \\
Ultrashort gamma-ray burst & U-GRB \\ 
White dwarf                & WD \\
X-ray flash                  & XRF \\
\hline
\end{tabular}
\caption{Alphabetic ordered list of the acronyms used in this work.}
\label{acronyms}
\end{table}

\section{Summary of the seven subclasses of GRBs}\label{sec:2}

\begin{table*}
\centering
\begin{tabular}{lcccccccc}
\hline
     & Sub-class  & Number & \emph{In-state}  & \emph{Out-state} & $E_{\rm p,i}$ &  $E_{\rm iso}$  &  $E_{\rm iso,Gev}$  \\
& & & & & (MeV) & (erg) &  (erg)\\  		
\hline

I   & S-GRFs & $18$ &NS-NS & MNS & $\sim0.2$--$2$ &  $\sim 10^{49}$--$10^{52}$  &  $-$ \\
II    & S-GRBs  & $6$ &NS-NS & BH & $\sim2$--$8$ &  $\sim 10^{52}$--$10^{53}$ &   $\gtrsim 10^{52}$\\

III    & XRFs & $49$ &CO$_{\rm core}$-NS    & $\nu$NS-NS & $\sim 0.004$--$0.3$  &  $\sim 10^{48}$--$10^{52}$ &    $-$ \\
IV   & BdHNe I  & $324$ &CO$_{\rm core}$-NS  & $\nu$NS-BH & $\sim0.2$--$2$ &  $\sim 10^{52}$--$10^{54}$ &    $\gtrsim 10^{52}$ \\
V  & BdHNe II  & $5$ &CO$_{\rm core}$-BH  & $\nu$NS-BH & $\gtrsim2$ &  $>10^{54}$ &   $\gtrsim 10^{53}$   \\
VI   & U-GRBs & $0$ &$\nu$NS-BH & BH & $\gtrsim2$ &  $>10^{52}$ & $-$ \\
VII  & GRFs  & $1$ &NS-WD & MNS & $\sim0.2$--$2$ &  $\sim 10^{51}$--$10^{52}$  & $-$\\
VIII  & GR-K  & $1$ &WD-WD & MWD & $\lesssim 0.1$ &  $\lesssim 10^{47}$  & $-$\\
\hline
\end{tabular}
\caption{Summary of the GRB subclasses. In addition to the subclass name, we report the number of GRBs for each subclass. We recall as well the ``in-state'' representing the progenitors and the ``out-state'' as well as the $ E_{\rm p,i}$ and $E_{\rm iso}$  for each subclass. We finally indicate the GeV emission in the last column which for the long GRBs is only for the {BdHNe I and BdHNe II (BH-SN)}, in the case of short bursts is only for S-GRBs and in all of them the GeV emission has energy more than $\sim 10^{52}$~erg.}
\label{tab:a}
\end{table*}

We address the specific role of the GeV radiation in order to further characterize the eight subclassess of GRBs presented in \citet{2016ApJ...832..136R} and {updated in} \citet{2018ApJ...852...53R} and here further updated. In Table~\ref{tab:a} we have indicated, for each GRB subclass, their name, the number of observed sources with  cosmological redshift, their progenitors characterizing their ``in-state''. In all cases  the progenitors are binary systems composed of various combinations of CO$_{\rm core}$, undergoing a SN explosion, of the $\nu$NS created in such SN explosion, of NS, of white dwarfs (WDs), of a BH. The ``out-state'' of the merging process, have been represented in Fig.~7 in \citet{2016ApJ...832..136R}.

We focus on the difference between XRFs and BdHNe.  

Two possible progenitor systems of long GRBs have been identified \citep{2016ApJ...832..136R}: a binary system composed of a CO$_{\rm core}$, exploding as SN Ic, and a NS companion \citep[see, e.g.,][]{Ruffini2001c,Rueda2012,2014ApJ...793L..36F}; and an analogous binary system where the binary companion of the exploding CO$_{\rm core}$ is an already formed BH. Binary X-ray sources such as Cygnus-X1 are possible progenitors of {these last} systems.

We address first the system with the NS binary companion. As the SN ejecta from the exploding CO$_{\rm core}$ engulfs the close NS binary companion \citep{2015ApJ...812..100B,2016ApJ...833..107B} a hypercritical accretion process occurs with the emission of $\nu\bar{\nu}$ pairs and, for tight binaries, the formation of an $e^+e^-$ plasma \citep{Rueda2012}. The presence of a NS companion explains the observed removing the outer layers of the CO$_{\rm core}$ \citep{2014ApJ...793L..36F}.

When the orbital period of the binary system is $\gtrsim 5$~min, the hypercritical accretion is not sufficient to trigger the collapse of the NS companion into a BH. Therefore, a MNS is formed creating a binary with the $\nu$NS originated in the SN explosion of the CO$_{\rm core}$. {The absence of the formation of the BH justifies their observed peak energy  in the range $4$~keV$<E_{\rm p,i}<300$~keV and isotropic energy in the range of $10^{48} \lesssim E_{\rm iso}\lesssim 10^{52}$~erg }  {and  have been indicated as  X-ray flash (XRF) in contrast with the more energetics long GRBs}\citep{2015ApJ...798...10R,2015ApJ...812..100B,2016ApJ...833..107B,2016ApJ...832..136R}.  

When the orbital period is as short as $\approx 5$~minutes, the hypercritical accretion proceeds at higher rates and the companion NS reaches its critical mass leading to 1) the formation of a BH in a binary system with the $\nu$NS \citep{2015PhRvL.115w1102F}, 2) the emission of a GRB with $E_{\rm iso}\gtrsim10^{52}$~erg and $E_{\rm p,i}\gtrsim0.2$~MeV, and 3) the onset of the GeV emission {when present is observed following the formation of }  the newly-born BH. These systems have been indicated as BdHNe \citep{2015ApJ...798...10R,2015ApJ...812..100B,2015PhRvL.115w1102F,2016ApJ...833..107B,2016ApJ...832..136R}. The BH formation and the associated GRB emission occur seconds after the onset of the SN explosion \citep[see, e.g., the case of GRB 090618 in][]{2012A&A...543A..10I}. 

The first list of BdHNe  was introduced in \citet{2016ApJ...833..159P} which was further extended in \citet{2018ApJ...852...53R}. 329 sources have been identified after the launch of \textit{Fermi} all with the given redshift, with their in-states represented by a CO$_{\rm core}$-NS binary and their out-state represented by a $\nu$NS, originated  in the SN explosion of a CO$_{\rm core}$ and a companion BH, their spectral peak energy has the range $0.2$~MeV$<E_{\rm p,i}<2$~MeV, isotropic energy $10^{52}<E_{\rm iso}<10^{54}$~erg and their isotropic GeV emission is $\sim 10^{52}$~erg.

Concerning the  {subclass with} progenitor a CO$_{\rm core}$-BH binary and their out-state represented by a $\nu$NS, originated in the SN explosion of a CO$_{\rm core}$ and a companion BH. Their spectral peak energy is larger than $2$~MeV, isotropic energy, $E_{\rm iso}>10^{54}$~erg and their isotropic GeV emission is $\sim 10^{53}$~erg.

{Such large values of the energetic has been justified on the ground of the previously existing BH or the formation of a newly born BH which prompts us to analyze in this article  the role of the GeV emission as a further confirmation of the process of the BH formation. }

\section{X-ray Flashes (XRFs)}\label{sec:3}

In our classification we have considered $48$ XRFs, long GRBs associated with SN with $E_{\rm iso}<10^{52}$~erg and $E_{\rm p,i}<300$~keV observed after the launch of \textit{Fermi} \citep{2018ApJ...852...53R}. {In our approach they}  originate from a progenitor of the binary CO$_{\rm core}$-NS in which due to large binary separation, typically more than $10^{11}$~cm, the critical mass of NS is not reached and the BH is not formed \citep{2016ApJ...832..136R}. {Our assumption that GeV radiation originated from BH leads to predict an absence of GeV radiation from XRFs}. 

{Only $7$ XRFs were observed by \textit{Fermi}: $3$ outside the boresight angle and $4$ inside, in none of them GeV emission has been observed see Table~\ref{tab:xrf}, offering a supporting  evidence linking the GeV emission to the rotational energy of the BH: no BH no GeV emission!}

\begin{table*}
\centering
\begin{tabular}{clccccccllll}
\hline\hline
Group 	          & XRF	    & $z$    &   	&	  $E_{\rm iso}$ 	&\vline		& XRF 		& $z$  		&&  $E_{\rm iso}$  &\\
	  	          &         &          & 	&	   ($10^{50}$~erg) &\vline  	&           &      		&  & ($10^{50}$~erg)&\\
\hline	          
	  	          &081007	&	$0.5295$	&				&	$17\pm2$		&\vline		&120224A	&	$1.1$	&				&	$37.2\pm7.6$&\\
                  &090407	&	$1.4485$	&				&	$50\pm17$		&\vline 	&120422A	&	$0.283$	&				&	$2.4\pm0.8$	&\\
                  &090417B	&	$0.345$		&				&	$30.9\pm2.7$	&\vline 	&120714B	&	$0.3984$&				&	$8.0\pm2.0$	&\\
                  &090814	&	$0.696$		&				&	$23.6\pm4.6$	&\vline		&120722A	&	$0.9586$&				&	$47.0\pm3.6$&\\
                  &091018	&	$0.971$		&				&	$81\pm10$		&\vline		&120724A	&	$1.48$	&				&	$58\pm13$	&\\
                  &100316B	&	$1.18$		&				&	$11.8\pm0.9$	&\vline 	&130511A	&	$1.3033$&				&	$16.7\pm5.3$&\\
                  &100316D	&	$0.059$		&				&	$0.59\pm0.05$	&\vline 	&130604A	&	$1.06$	&				&	$78\pm15$	&\\
                  &100418	&	$0.624$		&				&	$5.21\pm0.51$	&\vline 	&130702A	&	$0.145$	&				&	$6.5\pm1.0$	&\\
                  &100508A	&	$0.5201$	&				&	$8.7\pm1.4$		&\vline		&130831A	&	$0.4791$&				&	$46\pm2$	&\\
	  	          & 100724A &	$1.288$		&				&	$16.4\pm2.4$ 	&\vline		&131103A	&	$0.599$	&				&	$19.9\pm2.2$&\\
No \textit{Fermi} & 101219A &	$0.718$		&				&	$48.8\pm6.8$	&\vline		&140318A	&	$1.02$	&				&	$15.1\pm4.9$&\\
Observation 	  & 120804A &	$1.3$		&				&	$70.0\pm15.0$	&\vline		&140710A	&	$0.558$	&				&	$7.34\pm0.96$&\\
	  			  & 130603B &	$0.356$		&				&	$21.2\pm2.3$	&\vline		&150818A	&	$0.282$	&				&	$13.91\pm4.59$&\\
	  			  & 140622A &	$0.959$		&				&	$0.70\pm0.13$	&\vline		&150915A	&	$1.968$	&				&	$9.95\pm2.24$&\\
	  			  & 140903A &	$0.351$		&				&	$1.41\pm0.11$	&\vline		&151029A	&	$1.423$	&				&	$28.4\pm14.8$&\\
				  &101225A	&	$0.847$		&				&	$26\pm20$		&\vline 	&151031A	&	$1.167$	&				&	$76.3\pm7.1$&\\
				  &111005A	&	$0.013$		&				&	$0.0088\pm0.0016$&\vline	&160117B	&	$0.870$	&				&	$34.0\pm0.4$&\\
				  &111129A	&	$1.0796$	&				&	$52\pm12$		&\vline		&160314A	&	$0.726$	&				&	$32.0\pm4.6$&\\
				  &111225A	&	$0.297$		&				&	$3.55\pm0.56$	&\vline 	&160425A	&	$0.555$	&				&	$71.0\pm6.8$&\\
				  &111229A	&	$1.3805$	&				&	$9.0\pm1.2$		&\vline 	&161219B	&	$0.1475$&				&	$2.33\pm0.74$&\\
				  &120121B	&	$0.017$		&				&	$0.0139\pm0.0002$&\vline 	&			&	$$		&				&	$$&\\


			  
\hline\hline
Group		& XRF		& $z$       &	$E_{\rm p}$	& $E_{\rm iso}$     & Fermi GCN & $\theta$  & GeV observed   \\
			&         	&           &	(keV)			& ($10^{50}$~erg)   &           & (deg)     &           	& 		 & \\
\hline\hline
			&110106B	& $0.618$	& $209.851$		& $73\pm4$		& GCN 11543	& $103$	    & no	    	& 		 & \\
Outside		
Boresight 	&120907A	& $0.97$	& $304.365$		& $20.1\pm2.8$	& GCN 13721	& $110$	    & no   			& 		 & \\
Angle 		&130612A	& $2.006$	& $186.07$		& $71.7\pm7.6$	& GCN 14896 & $-$	    & no	    	& 		 & \\
\hline
Inside 		&101219B	& $0.55$	& $108.50$		& $63\pm6$ 		& GCN 11477	& $59$	    & no	    	&		 & \\	  
Boresight 	&121211A	& $1.023$	& $194.00$		& $61\pm22$		& GCN 14078	& $74$	    & no	    	& 		 & \\
Angle       &140606B	& $0.384$	& $654.63$		& $39.9\pm1.6$	& GCN 16363	& $66$	    & no	    	&		 & \\
            &150727A	& $0.313$	& $195.054$		& $19.52\pm2.56$& GCN 18081	& $46$	    & no	    	&		 & \\		 
\hline
\end{tabular}
\caption{\textit{{List of $48$  X-ray flashes (XRFs)}} divided in three different groups, the one  without \textit{Fermi}-LAT observation (upper group), 
with the \textit{Fermi}-LAT observation but the boresight  $\theta \geq 75^{\circ}$, the one within the \textit{Fermi}-LAT boresight angle (lower group, $\theta < 75^{\circ}$), none of the S-GRFs have any GeV photon detected. In the first column we indicate the name of the sources, in the second their redshift, in third column we indicate the $E_{\rm p,i}$ only deducible from the $Fermi$ data, in the fourth column we estimate $E_{\rm iso}$ which is systematically lower than the $10^{52}$~erg, we also add for convenience both the specific GCN of the $Fermi$ source as well as the boresight angle of the LAT observation in the column of the the non-observation of the GeV emission. The symbol ``$-$'' indicates no information on LAT boresight angle due to lack of GBM observation.}
\label{tab:xrf}
\end{table*}

\section{Binary-driven Hypernovae (BdHNe)}\label{sec:4}

We {now  address the $329$ BdHNe} with known redshift \citep{2018ApJ...852...53R}: out of them we are interested only in the  $21$ BdHNe with the boresight angle of \textit{Fermi}-LAT less than $75^{\circ}$ at the time of the trigger {and have as well TS value $>25$, which means to exclude at $5$--$\sigma$ the GeV photons from background sources.} Following the \textit{Fermi} catalog \citep{2013ApJS..209...11A} for time-resolved likelihood spectral analysis we divide the data into logarithmically spaced bins. If the TS value of each bin is smaller than $16$ we merge the time bin with the next one and we repeat the likelihood analysis. { In Table.~\ref{tab:cb}, in the first column we indicate the name of the BdHNe, in the second their redshift, in third column we present the $E_{\rm p,i}$ obtained from the \textit{ Fermi} data, in the fourth column we estimate $E_{\rm iso}$ which is larger than the $10^{52}$~erg, in the fifth column the \textit{ Fermi}  GCN numbers are shown, in sixth  column the lower limit values of $E_{\rm LAT}$ are provided and finally we add  boresight angle of the LAT and TS values of each GRBs observed by LAT. The values of $E_{\rm LAT}$ are calculated by multiplying the average luminosity in each time bin by the corresponding rest-frame time duration and then summing up all bins. We must point out that  at late time the GeV emission observation can  be prevented due to the instrument threshold of the LAT and is expected to give a minor contribution to the $E_{\rm LAT}$.}

\begin{table*}
\centering
\begin{tabular}{lccccccccl}
\hline\hline
Source        	&  $z$     &  $E_{\rm p,i}$   &  $E_{\rm iso}$    & Fermi GCN&  $E_{\rm LAT}$        &  $\theta$  &  TS    &   Comments    \\
        		&          &  (MeV)           &  ($10^{52}$~erg)  &			&  ($10^{52}$~erg)       &  (deg)     &        &                   \\
\hline
GRB 080916C	&  $4.35$  &  $2.27\pm0.13$   &  $407\pm86$	      &GCN $8246$&  $\gtrsim408\pm57$     &  $48.8$	&  $1450$  &   \\
GRB 090323A	&  $3.57$  &  $2.9\pm0.7$   &  $438\pm53$	      &GCN $9021$&  $\gtrsim48.85\pm0.59$ &  $57.2$	&  $150$   &   \\
GRB 090328A	&  $0.736$ &  $1.13\pm0.08$   &  $14.2\pm1.4$     &GCN $9044$&  $\gtrsim3.04\pm0.01$  &  $64.6$	&  $107$   &   \\
GRB 090902B	&  $1.822$ &  $2.19\pm0.03$   &  $292\pm29.2$     &GCN $9867$&  $\gtrsim110\pm5$	   	 &  $50.8$	&  $1832$  &   \\
GRB 090926A	&  $2.106$ &  $0.98\pm0.01$   &  $228\pm23$       &GCN $9934$&  $\gtrsim151\pm7$	     &  $48.1$  &  $1983$  &   \\
GRB 091003A	&  $0.897$ &  $0.92\pm0.04$   &  $10.7\pm1.8$	  &GCN $9985$&  $\gtrsim1.29\pm0.03$  &  $12.3$	&  $108$   &   \\
GRB 091208B	&  $1.063$ &  $0.25\pm0.04$   &  $2.10\pm0.11$	  &GCN $10266$&  $\gtrsim0.41\pm0$  &  $55.6$	&  $20$   &   \\
GRB 100414A	&  $1.368$ &  $1.61\pm0.07$   &  $55.0\pm0.5$	  &GCN $10594$	& $\gtrsim8.79\pm0.31$  &  $69$	&  $81$	   &   \\
GRB 100728A	&  $1.567$ &  $1.00\pm0.45$   &  $72.5\pm2.9$	  &GCN $11006$&  $\gtrsim1.15\pm0.2$  &  $59.9$	&  $32$	   &   \\
GRB 110731A	&  $2.83$  &  $1.21\pm0.04$   &  $49.5\pm4.9$	  &GCN $12221$&  $\gtrsim31.4\pm7.4$   &  $3.4$   &  $460$   &   \\
GRB 120624B 	&  $2.197$ &  $1.39\pm0.35$   &  $347\pm16$	      &GCN $13377$	&  $\gtrsim28\pm2$  &  $70.8$	&  $312$	   &   \\
GRB 130427A	&  $0.334$ &  $1.11\pm0.01$   &  $92\pm13$	      &GCN $14473$&  $\gtrsim5.69\pm0.05$  &  $47.3$  &  $163$   &   \\
GRB 130518A  	&  $2.488$ &  $1.43\pm0.38$   &  $193\pm1$	      &GCN $14675$	&  $\gtrsim3.5\pm0.6$  &  $41.5$	&  $50$	   &   \\
GRB 131108A  	&  $2.40$ &  $1.27\pm0.05$   &  $51.2\pm3.83$	  &GCN $15464$	&  $\gtrsim50.43\pm 5.86$  &  $23.78$	&  $870$	   &   \\
GRB 131231A	&  $0.642$ &  $0.27\pm0.01$   &	 $21.50\pm0.02$	  &GCN $15640$&  $\gtrsim2.18\pm0.02$  &  $38$	&  $110$   &   \\
GRB 141028A	&  $2.33$  &  $0.77\pm0.05$   &  $76.2\pm0.6$	  &GCN $16969$&  $\gtrsim7.36\pm0.46$  &  $27.5$	&  $104.5$ &   \\
GRB 150314A  	&  $1.758$ &  $0.86\pm0.01$   &  $70.1\pm3.25$	  &GCN $17576$	&  $\gtrsim1.93\pm0.89$  &  $47.13$	&  $27.1$	   &   \\
GRB 150403A  	&  $2.06$ &  $0.95\pm0.04$   &  $87.3\pm7.74$	  &GCN $17667$	&  $\gtrsim7.55\pm5.19$  &  $55.2$	&  $37$	   &   \\
GRB 150514A  	&  $0.807$ &  $0.13\pm0.01$   &  $1.14\pm0.03$	  &GCN $17816$	&  $\gtrsim0.42\pm0.05$  &  $38.5$	&  $33.9$	   &   \\
GRB 160509A	&  $1.17$  &  $0.80\pm0.02$   &  $84.5\pm2.3$	  &GCN $19403$&  $\gtrsim35.92\pm0.26$ &  $32$	&  $234$   &   \\
GRB 160625B	&  $1.406$  &  $1.3\pm0.1$   &  $337\pm 1$	  &GCN $19581, GCN 19604 $&  $\gtrsim 29.90\pm3.51$ &  $41.46$	&  $961.33$   &   \\ 
\hline
\end{tabular}
\caption{\textit{{List of 21 {long GRBs} inside Fermi-LAT boresight angle and GeV photon detected}}: the prompt and GeV emission properties of 21 {long GRBs} with GeV emission detected. Columns list: the source name, $z$, $E_{\rm p,i}$, $E_{\rm iso}$, $E_{\rm LAT}$, the position of the source from the LAT boresight $\theta$, the likelihood test statistic (TS). The $E_{\rm LAT}$  includes only the energy in the observed time duration, which does not cover the whole GeV emission period, and is different for each GRB, so we put a symbol '$\gtrsim$' to indicate the value is the lower limit. }
\label{tab:cb}
\end{table*}

Theoretically, {that the Kerr-Newman BH can have a fundamental role in the creation of $e^+e^-$ plasma responsible for acceleration of the prompt radiation has been known for some time \citep{1999A&AS..138..511R} and also a specific example in \cite{2017arXiv171205001R}. We are now turning to a different problem the possibility that the Kerr-Newman BH originates the GeV radiation. The above  two processes share some commonalities, for instance  there is a time-delay between  of the first GeV photon with respect to the trigger time of GRB as well as a similar delay in the onset of prompt phase  \citep[see, e.g.,][and Fig.~\ref{fig:02}]{2014MNRAS.443.3578N,2015ApJ...798...10R,2015ApJ...808..190R,2016ApJ...831..178R,2016ApJ...832..136R,2017ApJ...844...83A}, and this is the topic that we will addressed in further publication.}

{We recall that in  BdHNe there are two possibilities that the hypercritical accretion of the HN ejecta occurs on a NS companion leading to the formation of the BH (BdHN I) or it occurs on an already formed BH (BdHN II). We are going to see in the next sections how relating GeV radiation  to the loss of rotational energy of the Kerr-Newman can lead to the determination of the mass and spin of the BH in the above $21$ GRBs and create a separatix between BdHNe I and BdHNe II.}

\section{The determination of the mass and spin of the BH in BdHNe}\label{sec:5}

The first proposal for explaining  of the GeV energetics {as originating from the} the mass and spin of Kerr BH was introduced in  \citep{2016ApJ...831..178R,2016ApJ...832..136R,2017ApJ...844...83A, 2018arXiv180207552R} for short GRBs. We here apply their procedure for the case of BdHNe.

The extractable energy of a Kerr-Newman BH $E_{\rm extr}$ {is given by the subtracting the irreducible mass, $M_{irr}$, from  the total mass of the BH, $M$:}
\begin{equation}
\begin{split}
\label{EextrFull}
E_{extr}= M-M_{irr}\\
=M-\sqrt{\dfrac{M^2-Q^2/2+M\sqrt{M^2-Q^2-a^2}}{2}} \\
=M\Bigg(1-\sqrt{\dfrac{1-\lambda^2/2+\sqrt{1-\lambda^2-\alpha^2}}{2}}
\Bigg),
\end{split}
\end{equation}
{where we have used the BH mass-formula (\ref{aone}) and introduced the dimensionless charge and angular momentum BH parameters, $\lambda=Q/M$ and $\alpha=a/M=J/M^2$, respectively.}

{In the following we assume that the electric charge contribution to the geometry is negligible, i.e.~$\lambda \ll 1$ . The electric charge is nevertheless essential in  contributing the electromagnetic process and in producing the GeV radiation. We shall return in this topic  in a forthcoming article. Here the main goal is to show that the rotational energy of the Kerr-Newman BH is indeed sufficient to explain the energetic of the GeV emission and, in turn it leads to determine the mass and spin of BH.}

{By neglecting the electromagnetic term, the extractable energy becomes}
\begin{equation}
\label{Eextr}
E_{\rm extr}=M-M_{\rm irr}=\left(1-\sqrt{\frac{1+\sqrt{1-\alpha^2}}{2}}\right)M,
\end{equation}
which we use  to obtain $M$ as a function of $\alpha$, $M(\alpha)$, by requesting the condition that observed GeV emission originates from BH extractable energy, i.e.
\begin{equation}
\label{latextr}
E_{\rm LAT} = E_{\rm extr}.
\end{equation}

{Equation (\ref{Eextr}) has two  parameters $M$ and $\alpha$, we therefore need another equation to determine both $M$ and $\alpha$.}

{In the case of BdHNe I, the BH  originates from the hypercritical accretion of the SN on the NS, the formation of the BH occurs when the NS reaches its critical mass and the mass of the BH satisfies the condition}
\begin{equation}
\label{meq}
M\geq M_{\rm crit}(\alpha).
\end{equation}

Concerning the NS critical mass value, it has been shown in \citet{2015PhRvD..92b3007C} that for the NL3, GM1 and TM1 EOS it is fitted, with a maximum error of $0.45\%$, by the relation
\begin{equation}\label{eq:Mcrit}
M_{\rm crit}(\alpha)=M_{\rm crit}^{J=0}(1 + k j^p),
\end{equation}
where $k$ and $p$ are parameters that depend upon the nuclear EOS, $M_{\rm crit}^{J=0}$ is the critical mass in the non-rotating (spinless) case and $j$ is an angular momentum parameter related to $\alpha$ by
\begin{equation}
j \equiv \alpha (M_{\rm crit}/M_\odot)^2.
\end{equation}

\begin{figure}
\centering
\includegraphics[width=\hsize,clip]{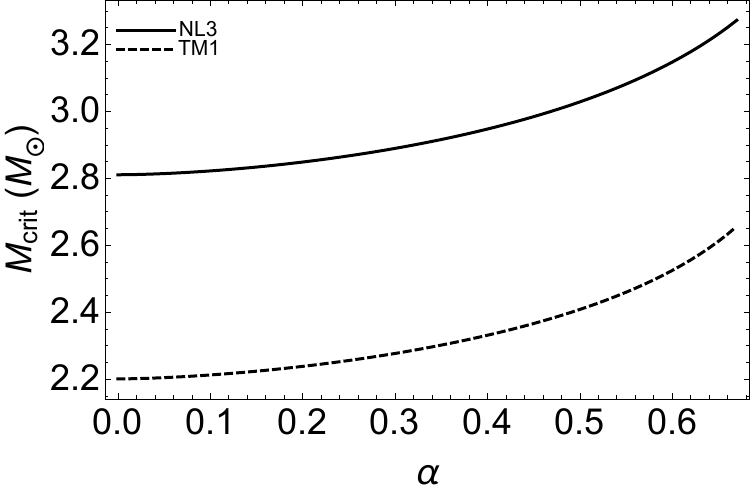}
\caption{NS critical mass as a function of the spin parameter $\alpha$ for the NL3 and TM1 EOS. We recall that the maximum spin parameter of a uniformly rotating NS is $\alpha_{\rm max}\approx0.71$, independently of the NS EOS; see e.g. \citet{2015PhRvD..92b3007C}.}
\label{fig:Mcrit}
\end{figure}

{Therefore, Eq.~(\ref{eq:Mcrit}) is an implicit equation of the NS critical mass as a function of the spin parameter $\alpha$ which can be solved numerically. We show in Fig.~\ref{fig:Mcrit} such a relation for the NL3 and TM1 EOS. The maximum spin parameter is independent of the EOS of a uniformly rotating NS and is $\alpha_{max}\approx 0.7$. For more details  see \citet{2015PhRvD..92b3007C} and \citet{2018arXiv180207552R}.} {As it can be seen in Fig.~\ref{fig:Mcrit} considering two EOS of NS, for each value of $E_{\rm extr}$ there are two corresponding solution for $M(\alpha)$ and $\alpha$ which are listed in Table~\ref{tab:dd}.}

In $16$ {of the above long GRBs in Table~\ref{tab:cb}}, the minimum BH mass, and corresponding maximum BH spin parameter are respectively in the range of $2.21\leq M\leq 2.64~M_\odot$ and $0.05\leq\alpha\leq 0.71$, for the TM1 model, and $2.81\leq M\leq 3.23~M_\odot$ and $0.04\leq\alpha\leq 0.65$, for the NL3 model. 

{We can then conclude that these GRBs are BdHNe with an ``in-state'' CO$_{\rm core}$-NS and ``out-state'' a $\nu$NS and a newly formed BH close to the critical mass of the NS, we are here indicated this family as BdHNe type one, BdHNe I. We now turn to the remaining $5$ sources in Table~\ref{tab:cb} and show that indeed they belong to BdHNe originating from hypercritical accretion of the HN on an already formed BH which will indicate as BdHNe type two, BdHNe II.}

\begin{table*}
\centering
\begin{tabular}{lcccc}
\hline\hline
                &  \multicolumn{2}{c}{TM1}                                             &  \multicolumn{2}{c}{NL3} \\
\cline{2-5}
Source          &  $\alpha$                        &   $M(\alpha)$                     &  $\alpha$                           &   $M(\alpha)$ \\
        	&                                  &   (M$_\odot$)                     &                                     &   (M$_\odot$) \\

BdHN 090328A	&  $0.2434^{+0.0004}_{-0.0004}$    &  $2.2526^{+0.0001}_{-0.0001}$     &  $0.2167^{+0.0003}_{-0.0003}$       &  $2.8538^{+0.0001}_{-0.0001}$ \\
BdHN 091003A	&  $0.161^{0.002}_{-0.002}$         &  $2.2259^{0.0005}_{-0.0005}$       &  $0.143^{+0.001}_{-0.001}$          &  $2.8311^{+0.0004}_{-0.0004}$ \\
BdHN 091208B	&  $\lesssim 0.0910$         &  $\gtrsim 2.2101$       &  $\lesssim 0.0806$          &  $\gtrsim2.818$ \\
BdHN 100414A	&  $0.400^{+0.006}_{-0.006}$       &  $2.330^{+0.004}_{-0.004}$        &  $0.359^{+0.006}_{-0.006}$          &  $2.921^{+0.003}_{-0.003}$ \\
BdHN 100728A	&  $0.1516^{+0.0124}_{-0.0136}$    &  $2.2235^{+0.0033}_{-0.0034}$     &  $0.1345^{+0.0111}_{-0.0121}$       &  $2.8291^{+0.0028}_{-0.00286}$ \\
BdHN 110731A	&  $0.68^{+0.05}_{-0.06}$          &  $2.57^{+0.12}_{-0.10}$           &  $0.62^{+0.05}_{-0.06}$             &  $3.18^{+0.09}_{-0.09}$ \\
BdHN 120624B	&  $0.6429^{+0.0154}_{-0.0167}$    &  $2.5966^{+0.0320}_{-0.0307}$     &  $0.5930^{+0.0160}_{-0.0170}$       &  $3.1362^{+0.0236}_{-0.0232}$ \\
BdHN 130427A	&  $0.327^{+0.001}_{-0.001}$        &  $2.2893^{+0.0006}_{-0.0006}$    &  $0.293^{+0.001}_{-0.001}$          &  $2.8854^{+0.0005}_{-0.0005}$ \\
BdHN 130518A	&  $0.2604^{+0.0204}_{-0.0224}$     &  $2.2592^{+0.0084}_{-0.0086}$    &  $0.2320^{+0.0184}_{-0.0202}$       &  $2.8594^{+0.0073}_{-0.0074}$ \\
BdHN 131231A	&  $0.2075^{+0.0007}_{-0.0007}$    &  $2.2399^{+0.0002}_{-0.0002}$     &  $0.1844^{+0.0006}_{-0.0006}$       &  $2.8430^{+0.0002}_{-0.0002}$ \\	
BdHN 141028A	&  $0.37^{+0.01}_{-0.01}$          &  $2.312^{+0.006}_{-0.006}$        &  $0.331^{+0.009}_{-0.010}$          &  $2.905^{+0.005}_{-0.005}$ \\
BdHN 150314A	&  $0.1954^{+0.0395}_{-0.0511}$      &  $2.2360^{+0.0133}_{-0.0144}$   &  $0.1736^{+0.0354}_{-0.0456}$       &  $2.8397^{+0.0114}_{-0.0122}$ \\
BdHN 150403A	&  $0.3734^{+0.0976}_{-0.1580}$     &  $2.3141^{+0.0679}_{-0.0716}$    &  $0.3348^{+0.0907}_{-0.1432}$       &  $2.9067^{+0.0578}_{-0.0615}$ \\
BdHN 150514A	&  $0.0921^{+0.0053}_{-0.0056}$     &  $2.2103^{+0.0010}_{-0.0010}$    &  $0.0816^{+0.0047}_{-0.0050}$       &  $2.8181^{+0.0008}_{-0.0008}$ \\
BdHN 160509A	&  $0.707^{+0.002}_{-0.002}$       &  $2.636^{+0.004}_{-0.004}$        &  $0.651^{+0.002}_{-0.002}$          &  $3.232^{+0.003}_{-0.003}$ \\
BdHN 160625B	&  $0.6576^{+0.0244}_{-0.0280}$       &  $2.6270^{+0.0598}_{-0.0552}$        &  $0.6082^{+0.0259}_{-0.0288}$   &  $3.1586^{+0.0424}_{-0.0412}$ \\
\hline
\end{tabular}
\caption{The BH spin parameter $\alpha$ and mass $M$ within the TM1 and the NL3 nuclear models, as inferred from the values of $E_{\rm LAT}$ for $16$ BdHNe I, out of the $21$ long GRBs in Table.~\ref{tab:cb}, providing BH spin parameters $\alpha<0.71$, consistent with the maximum spin parameter of a rotating NS \citep[see][for details]{2015PhRvD..92b3007C}.}
\label{tab:dd}
\end{table*}

\section{Black hole \& hypernova binary (BdHNe II)}\label{sec:6}

In $5$ {long GRBs} out of the $21$ ones in Table~\ref{tab:cb} {the BH mass is always larger than the NS critical mass, i.e. $M > M_{\rm crit}(\alpha)$, even assuming the maximum value of the NS spin parameter, $\alpha_{\rm max}\geq$ 0.71}. Therefore, for these $5$ GRBs we fix $\alpha = \alpha_{\rm max}$ and compute the BH mass $M$ necessary to explain the values $E_{\rm LAT}$ in Table.~\ref{tab:cb}. These minimum masses are listed in the Table~\ref{tab:2b} and range in the interval $3.55\leq M\leq 29.7~M_\odot$. Such values are higher than those attainable in BdHNe I where the NS binary companion collapses into a BH with a $M \sim M_{\rm crit}(\alpha)$  {(see Fig.~\ref{fig:Mcrit})}.

{These sources evidence the possibility of a}
progenitor system {composed of} a CO$_{\rm core}$-BH binary instead of a CO$_{\rm core}$-NS binary. This possibility has been already advanced in \citet{2016ApJ...832..136R} and \citet{2016arXiv160203545R} and finds here a further confirmation.

{We can then conclude in generality that there are two types of BdHNe and most fundamental from an astrophysical point of view is that the rotational energy and spin of the BH are sufficient in both types, in fulfilling the energy requirement in GeV radiation in all observed above $21$ sources. In all observed above: $16$ BdHNe I $+$ $5$ BdHNe II $=$ $21$ BdHNe.}

\begin{table*}
\centering
\begin{tabular}{lcc}
\hline\hline
Source        	&  $\alpha$    &   $M(\alpha)$ \\
        	&              &   (M$_\odot$) \\
\hline
BdHN II 080916C	&  $0.71$      &  $29.7\pm3.4$  \\
BdHN II 090323A	&  $0.71$      &  $3.55\pm0.04$ \\
BdHN II 090902B	&  $0.71$      &  $7.99\pm0.36$ \\
BdHN II 090926A	&  $0.71$      &  $10.98\pm0.51$ \\
BdHN II 131108A 	&  $0.71$      &  $3.67\pm0.43$  \\
\hline
\end{tabular}
\caption{The $5$ BdHNe II, out of the $21$ long GRBs in Table~\ref{tab:cb}, requiring $M>M_{\rm max}^{J\neq0}$. The masses $M$ have been obtained from Eq.~4 in \citet{2018arXiv180207552R} by fixing $\alpha=\alpha_{\rm max}=0.71$. {The inferred values for mass of each system is bigger than the maximum critical-mass of the NS, for TM1 model $M_{crit}(\alpha_{max})=2.62 M_{\odot}$ and for NL3 model $M_{crit}(\alpha_{max})=3.38 M_{\odot}$, see Fig.~\ref{fig:Mcrit}. These sources cannot be associated with NS, therefore we conclude the hypercritical accretion occurs to an already formed.}}
\label{tab:2b}
\end{table*}

\section{BdHNe without GeV emission and Geometry of the BdHNe}\label{sec:10}

{Initially} unexpected it has been  the presence of $27$ BdHNe within the LAT boresight, {without the} GeV emission, see Table~\ref{tab:BdHNe_No_GeV}. Although the distribution of the {boresight} angle and redshift, is totally analogous to the {one of the $21$ sources considered in sections~\ref{sec:4}, \ref{sec:5} and \ref{sec:6}} but no GeV emission is observed. {Some BdHNe with \textit{Swift} data have been identified as sources of hard and soft X-ray flares  as well as of the extended thermal emission, \citet[see details in ][]{2018ApJ...852...53R}. A complete example has been given for GRB 151027A by \citet{2017A&A...598A..23N} and \citet{2017arXiv171205001R}. There we assumed that the viewing angle of these sources lies in the equatorial plane of the progenitor system. We here assume that the 21 sources consider in Table~\ref{tab:cb} have an viewing angle seen from the normal of the plane, while the remaining 27 have a viewing angle in the equatorial plane. This allows us to introduce a new morphology for the BdHNe and predict specific observational properties.}

\begin{table*}
\centering
\begin{tabular}{lccclccl}
\hline\hline
BdHNe	& $z$       &	$E_{\rm p}$	        & $E_{\rm iso}$     & Fermi GCN     & $\theta$  & GeV observed & comments \\
        &           &	(MeV)			        & ($10^{52}$~erg)   &               & (deg)     &               & \\
\hline	          
081222  &   $2.77$  &   $0.51\pm0.03$                  &   $27.4\pm2.7$    &   GCN 8715    &   $50.0$  &   no      & \\
090424A &   $0.544$ &   $0.27\pm 0.04$                  &   $4.07\pm0.41$   &   GCN 9230    &   $71.0$  &   no      & \\
090516A	&	$4.109$	&	$0.14\pm0.03$	&	$99.6\pm16.7$	&	GCN 9415	&	$20.0$	&	no	    & Clear X-ray flare observed\\
091127A &   $0.49$  &   $0.05\pm 0.01$                  &   $1.64\pm0.18$   &   GCN 10204   &   $25.0$  &   no      & \\
100615A &   $1.398$ &   $0.21\pm 0.02$                  &   $5.81\pm0.11$   &   GCN 10851   &   $64.0$  &   no      & \\
100728B &   $2.106$ &   $0.32\pm 0.04$                  &   $3.55\pm0.36$   &   GCN 11015   &   $57.1$  &   no      & \\
110128A &   $2.339$ &   $0.46\pm 0.01$                    &   $1.58\pm0.21$   &   GCN 11628   &   $45.0$  &   no      & \\
111228A &   $0.716$ &   $0.060\pm0.007$                  &   $2.75\pm0.28$   &   GCN 12744   &   $70.0$  &   no      & \\
120119A &   $1.728$ &   $0.52\pm0.02$                  &   $27.2\pm3.6$    &   GCN 12874   &   $31.4$  &   no      & \\
120712A &   $4.175$ &   $0.64\pm 0.13$                  &   $21.2\pm2.1$    &   GCN 13469   &   $42.0$  &   no      & \\
120716A &   $2.486$ &   $0.4\pm 0.04$                  &   $30.2\pm3.0$    &   GCN 13498   &   $63.0$  &   no      & \\
120909A &   $3.93$  &   $0.87\pm 0.01$                   &   $87\pm10$       &   GCN 13737   &   $66.0$  &   no      & \\
130528A &	$1.250$	&	$0.27\pm0.18$	&	$18.01\pm2.28$	&	GCN 14729	&	$60.0$	&	no	    & X-ray flare observed\\
130925A &	$0.347$	&	$0.14\pm0.04$		&	$3.23\pm0.37$	&	GCN 15261	&	$22.0$	&	no	    & X-ray flare observed\\
131105A &   $1.686$ &   $0.55\pm0.08$       &   $34.7\pm1.2$    &   GCN 15455   &   $37.0$  &   no      & \\
140206A&   $2.73$  &   $1.1\pm0.03$	&	$144.24\pm19.20$&   GCN 15790   &   $46.0$  &   no   & Clear X-ray flare observed\\
140213A &   $1.2076$&   $0.176\pm0.004$       &   $9.93\pm0.15$   &   GCN 15833   &   $48.5$  &   no      & \\
140423A &   $3.26$  &   $0.53\pm0.04$       &   $65.3\pm3.3$    &   GCN 16152   &   $44.0$  &   no      & \\
140623A &   $1.92$  &   $1.02\pm0.64$        &   $7.69\pm0.68$   &   GCN 16450   &   $32.0$  &   no      & \\
140703A &   $4.13$  &   $0.91\pm0.07$        &   $1.72\pm0.09$   &   GCN 16512   &   $16.0$  &   no      & \\
140907A &   $1.21$  &   $0.25\pm0.02$	&	$2.29\pm0.08$   &   GCN 16798   &   $16.0$  &   no      & X-ray flare observed\\
141220A &   $1.3195$&   $0.42\pm0.02$       &   $2.44\pm0.07$   &   GCN 17205   &   $47.0$  &   no      & \\
150301B &   $1.5169$&   $0.45\pm0.10$   &   $2.87\pm0.42$   &   GCN 17525   &   $39.0$  &   no      & \\
150821A &   $0.755$ &   $0.57\pm0.03$   &   $14.7\pm1.1$    &   GCN 18190   &   $57.0$  &   no      & \\
151027A &   $0.81$  &   $0.62\pm0.11$	&	$3.94\pm1.33$   &   GCN 18492   &   $10.0$  &   no      & Clear X-ray flare observed\\
151111A &   $3.5$   &   $0.25\pm0.04$	&	$3.43\pm1.19$   &   GCN 18582   &   $50.0$  &   no      & X-ray flare observed\\
161014A &   $2.823$ &   $0.64\pm0.06$   &   $10.1\pm1.7$    &   GCN 20051   &   $69.0$  &   no      & \\
\hline
\end{tabular}
\caption{\textit{{List of $27$ BdHNe inside Fermi-LAT boresight angle and no GeV photon detected}}: $27$ BdHNe with redshift taken from \citep{2016ApJ...832..136R} from 2008, when \textit{Fermi} started to operate, till the end of 2016. All of them are within the boresight of Fermi-LAT, but none detected GeV photon. For each source the columns list: $z$, $E_{\rm iso}$, $E_{\rm p}$, GCN number, position of the source from LAT boresight $\theta$, whether was detection by LAT, and additional information.}
\label{tab:BdHNe_No_GeV}
\end{table*}

 We look at the ratio between the number of GRBs emitting GeV radiation $N_{\rm LAT}$, and the total number of GRBs $N_{\rm tot}$ within the LAT FoV. Assuming a two-side cone emitting region, the half-opening angle of a single cone $\vartheta$ is given by
\begin{equation}
\label{angle}
1-\cos\vartheta = \frac{N_{\rm LAT}}{N_{\rm tot}}.
\end{equation}
\begin{figure*}
\centering
\includegraphics[width=0.49\hsize,clip]{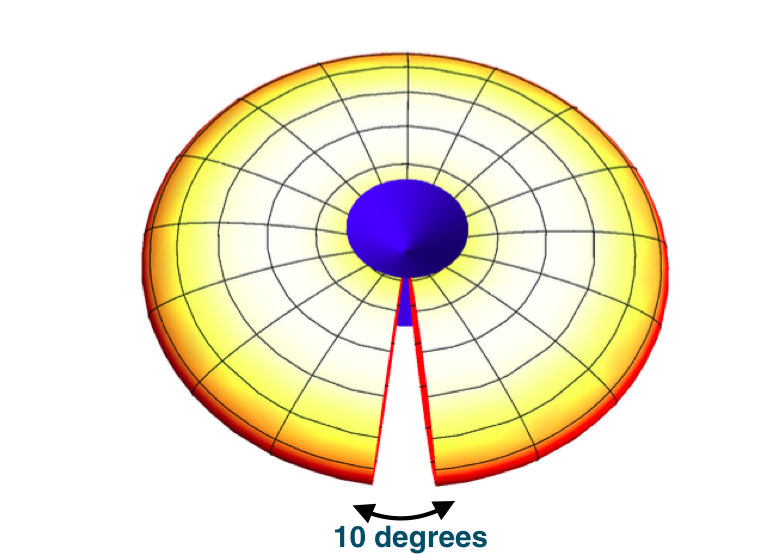}
\includegraphics[width=0.49\hsize,clip]{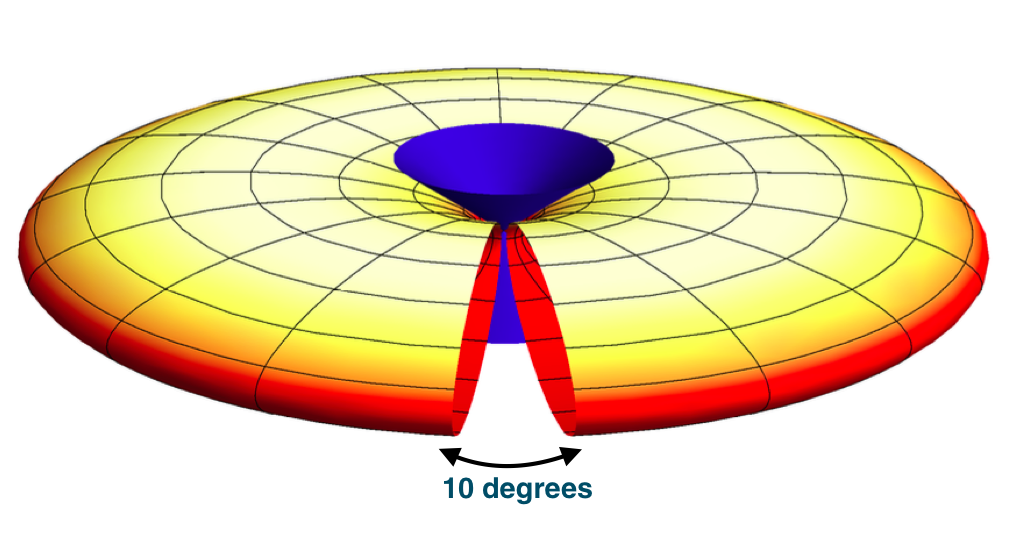}
\caption{Idealized plot for showing the morphology of the BdHNe. The GeV emission is detectable when the viewing angle is less than the $55.77^{\circ}$ from the normal to the orbital plane. Left panel is the situation in which the detectors can observe the GeV emission and the right panel is the one which GeV emission is not detectable and only Gamma-ray and X-ray flares are detectable.{The $10^{\circ}$ cuts in both figures indicate the low density region in Fig~\ref{fig:cc} through which the prompt radiation phase can be observed in the sources with viewing angle in the equatorial plane.}}
\label{fig:dd}
\end{figure*}

Our search in the LAT data\footnote{\url{https://fermi.gsfc.nasa.gov/ssc/observations/types/grbs/lat_grbs/table.php}} gives $N_{\rm LAT}=21$ and $N_{\rm tot}=48$, leading to $\vartheta=55.77^\circ \approx 60^\circ$. Therefore, in BdHNe the GeV emission comes from a wide angle emission as can be seen in idealized Fig.~\ref{fig:dd}. 

Therefore, the following general conclusion can be reached: a {new} morphology {of the BdHN is identifiable confirmed by the GeV emission in this paper, the soft and hard X-ray flares in \citet{2018ApJ...852...53R}, the extended thermal emission in \citet{2017A&A...598A..23N} and guided by} the large number of numerical simulations describing the accretion of the SN ejected material around the NS companion (see Fig.~\ref{fig:SPHsimulation} and Fig.~\ref{fig:cc}) and its idealized representation in Fig.~\ref{fig:dd}.

What can be concluded from the above results is that in BdHNe, in view of the ejected SN accreting material, the GeV emission is only detectable when the viewing angle is less than the $\approx 60^\circ$ from the normal to the plane, see left plot in Fig.~\ref{fig:dd}. Whenever the viewing angle is within $60^\circ$ from the co-planarity condition no GeV emission is observed though X-ray and Gamma-ray Flares are observed, see right plot in Fig.~\ref{fig:dd}. 

\section{SPH simulation of BdHNe}\label{sec:11}

\begin{figure*}
\centering
\includegraphics[width=0.85\hsize,clip]{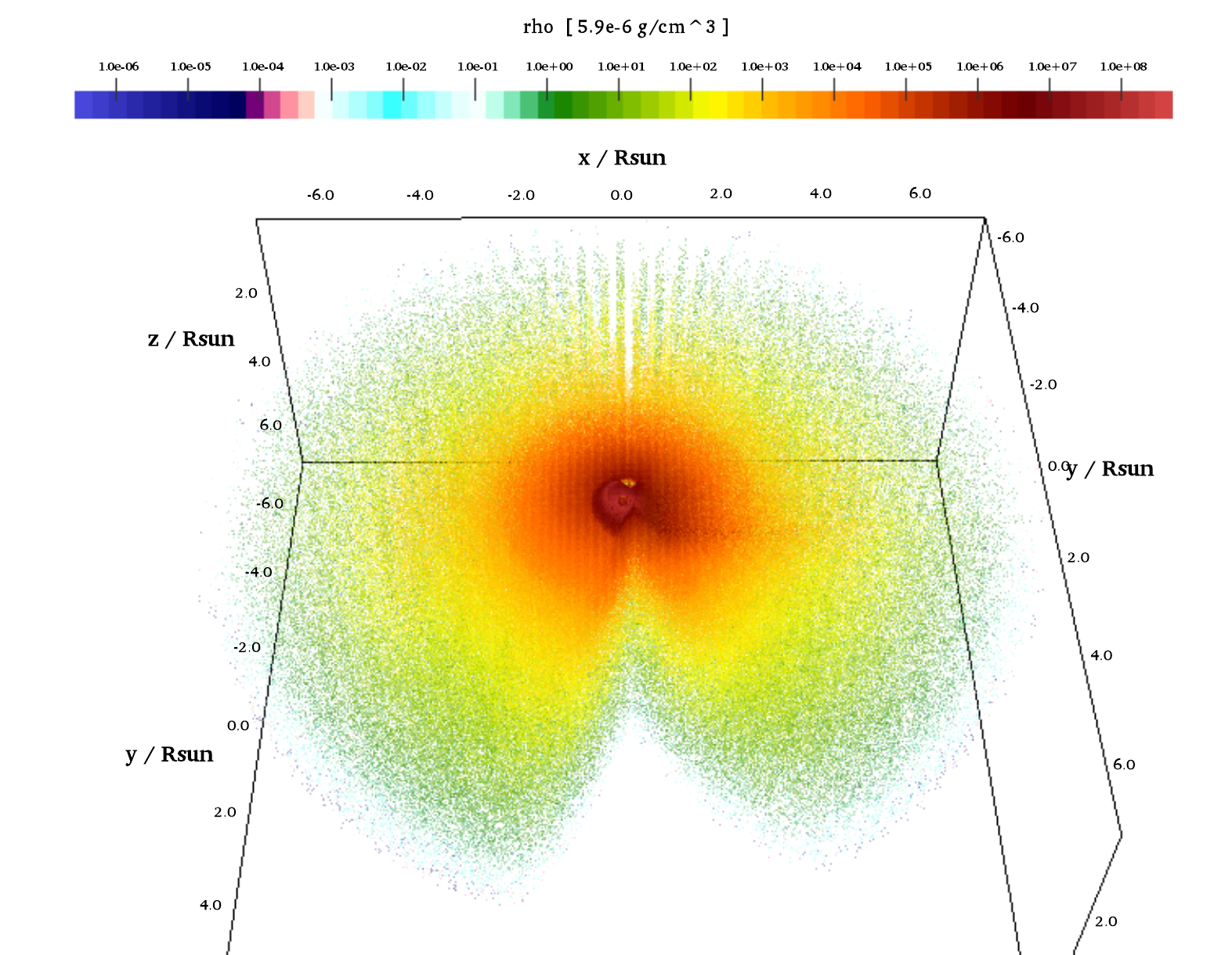}
\caption{Three-dimensional, half hemisphere views of the density distribution of the SN ejecta at the moment of BH formation in a BdHN. The simulation is performed with a SPH code that follows the SN ejecta expansion under the influence of the NS companion gravitational field including the effects of the orbital motion and the changes in the NS gravitational mass by the hypercritical accretion process. The initial conditions of the SN ejecta are set by a homologous velocity distribution in free expansion and the mass-distribution is modeled with $16$ millions point-like particles \citep[see][for additional details]{2016ApJ...833..107B}. The binary parameters of this simulation are: the NS companion has an initial mass of $2.0~M_\odot$; the CO$_{\rm core}$, obtained from a progenitor with zero-age main-sequence (ZAMS) mass $M_{\rm ZAMS}=30~M_\odot$, leads to a total ejecta mass $7.94~M_\odot$ and to a $1.5~M_\odot$ $\nu$NS, the orbital period is $P\approx 5$~min (binary separation $a\approx 1.5\times 10^{10}$~cm). {The distribution of the ejecta is not axially symmetric; it is strongly influenced by the rotation of the system and accretion occurring in the binary component, see Fig.~\ref{fig:SPHsimulation}. Particularly relevant for the observations is the low density region of $\approx 10^{\circ}$ which allows, for the sources with viewing angle in the equatorial plane to detect the prompt radiation phase. This has been qualitatively indicated in Fig.~\ref{fig:dd}. In this sources only a fraction of approximately 10$\%$ the prompt radiation can be detectable, they are the only ones able to trigger the \textit{Fermi}-GBM and the remaining 90$\%$ will not have the prompt radiation detectable, see \citet{2017arXiv171205001R} and work in progress.}}
\label{fig:cc}
\end{figure*}

The numerical simulation at the moment of BH formation in a BdHN is presented in \citet{2016ApJ...833..107B,2018arXiv180304356B}. Three-dimensional (3D) views of the density distribution at the moment of the BH formation in a BdHN are shown Fig.~\ref{fig:cc}. These plots correspond to the simulation of the SN ejecta expansion under the presence of the NS companion. The simulation is performed with a SPH technique in which the SN ejecta material is divided by $N$ point-like particles, in the present case 16 million, with different masses and followed their motion under the NS gravitational field. The orbital motion of the NS around the SN explosion center is taking into account as well as the NS star gravitational mass changes via the hypercritical accretion process. The last was modeled independently estimating the accretion rate on the NS with the Bondi-Hoyle formalism. For the initial conditions of the simulation was adopted a homologous velocity distribution in free expansion and the power law initial density profile of the SN matter was modeled by populating the inner layers with more particles \citep[see][for additional details]{2016ApJ...833..107B,2018arXiv180304356B}. Figures~\ref{fig:SPHsimulation} and~\ref{fig:cc} corresponds to an initial binary system formed by a $2\,M_\odot$ NS and the CO$_{\rm core}$, obtained from a progenitor with $M_{\rm ZAMS}=30\,M_\odot$. When the CO$_{\rm core}$ collapses and explodes, ejects $7.94\, M_\odot$ and leads a $\nu$NS of $1.5\,M_\odot$. The initial binary period is about $5$~min ($\approx 1.5\times 10^{10}$~cm). 

{The new morphology of the BdHNe presented here and in the  previous section, leads to a difference in the observed energy spectra and time variabilities for sources with viewing angle in the plane or normal to the orbital plane of the binary progenitor. We infer that our $21$ BdHNe, with viewing angle less than $ \approx 60^{\circ}$ from the normal to the orbital plane of the binary progenitor, ``seen from above'', have larger  $E_{\rm iso}$ than the ones with the viewing angle lying in the plane of the binary system (see Tables~\ref{tab:cb} and \ref{tab:BdHNe_No_GeV}). This explains the association/non-association of the GeV emission with bright GRBs often mentioned in the current literature (see \citealp{2011ApJ...738..138R,2011ApJ...732...29C} and Fig.~4 in \citealp{2018arXiv180401524N}).}

{An additional issue in the traditional approach (see e.g.~\citealp{2015MNRAS.454.1073B, 2011ApJ...738..138R} and sections 3 and 4 in \citealp{2018arXiv180401524N}) is also solvable: the sources,  which are seen with a viewing angle lying in the orbital plane have stronger flaring activities in the X-ray afterglow compared to the $21$ emitting GeV, therefore the ratio between $E_{iso}$ and the luminosity in the X-ray afterglow is systematically smaller than in the $21$ with GeV emission. This offers a different explanation than the one presented in the traditional approach. However, all these matters as already mentioned in \citet{2017arXiv171205001R} need a new operational definition of $E_{\rm iso}$, taking into due account the hard and soft X-ray flares and the extended thermal emission.}

{
Another important, specific feature of the new morphology of BdHNe is the presence of the $\nu$NS formed at the center of the exploding SN (see Fig.~\ref{fig:SPHsimulation} and \citealp{2018arXiv180304356B,2016ApJ...833..107B}). We have shown that the $\nu$NS manifests through the synchrotron emission by relativistic electrons injected from it into the expanding magnetized SN ejecta, as well as through its pulsar emission which explain the early and late optical and X-ray afterglow, respectively, allowing the inference of the $\nu$NS rotation period (see \citealp{2017arXiv171205000R}). A smoking gun of this picture, namely the verification of the $\nu$NS activity following the above mechanism, both in XRFs and in BdHNe, and the connection of the inferred rotation period of the $\nu$NS to the one of the CO$_{\rm core}$ and to the orbital period, from angular momentum conservation, has been explicitly shown in the GRB 180728A (XRF) and GRB 130427A (BdHN); see Y.~Wang, et al., in preparation.
}

\section{The Ultra-relativistic GeV Emission}\label{sec:7}

{Having fulfilled the energy requirement of the GeV emission from the rotational energy of the Kerr-Newman BH, we can turn to the evaluation of the Lorentz factor of the GeV directly by the condition of transparency.}

Taking the GeV emission from a typical GRB, which has a power-law spectrum of  flux density with index $\sim -2$, namely
\begin{equation}
f = f_0 \epsilon^{-2},
\end{equation}
where $f_0$ is a constant from the fitting of the data.  Following Appendix \ref{apd:transparency}, the lower limit of the Lorentz factor from the $\gamma+\gamma \rightarrow e^{+}e^{-}$ suppression of the ultra-high energy photons gives
\begin{eqnarray}
\Gamma &=& 1525\,\left(\frac{f_0}{10^{-4}}\right)^{1/4}\left(\frac{D_L}{10^{27}\,{\rm cm}}\right)^{1/2}\left(\frac{\epsilon_{max}}{10~{\rm GeV}}\right)^{1/4}\nonumber\\ &&\times \left(\frac{t}{10~{\rm s}}\right)^{-1/4},
\end{eqnarray}
where $D_L$ is the luminosity distance, the value of parameters $f_0=10^{-4}$, $D_L=10^{27}$~cm, $t=10$~s, and $\epsilon_{max}=10$~GeV correspond to luminosity $L\sim 10^{52}$~erg~s$^{-1}$ at $10$~s and the highest energy photon observed is $\sim 10$~GeV.

\begin{figure}
\centering
\includegraphics[width=0.45\textwidth]{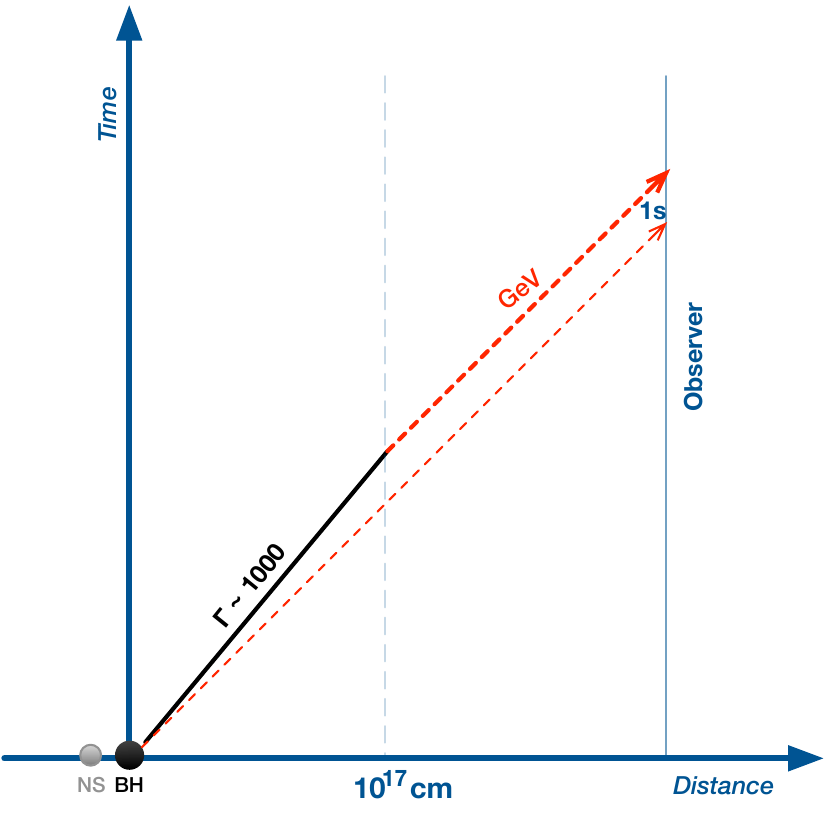} 
\caption{{The qualitative space-time diagram of the transparency point of the source of the GeV radiation when measured in the laboratory frame, see \citet{Ruffini2001a}. The black solid line represents the ultra-relativistic motion of the source at $\Gamma = 10^3$, the thin (thick) red dashed lines are the photon trajectories of the prompt (GeV) radiation. The emission of the GeV radiation occurs at $r=\Delta t_a c \sqrt{\Gamma^2-1} (\Gamma+\sqrt{\Gamma^2-1})\approx 2 \Gamma^2 c \Delta t_a\approx 10^{17}$~cm, where $\Delta t_a$ is the arrival time delay between the prompt and the GeV emission.}}
\label{Gevgamma}
\end{figure}

{
The independent origin of the GeV radiation as originating from the BH is further confirmed by the space-time diagram see Fig. \ref{Gevgamma}, illustrating how the condition of transparency for a $\Gamma \sim 1500$ occurs at $\approx 10^{17}$~cm. This fact introduces separatix between the ultra-relativistic emission from the BH and the plateau and the afterglow phase of the soft X-ray, which has been proved to be mildly-relativistic ($\Gamma \sim 2$) from the observation of the thermal component following the prompt radiation phase \citep{2015ApJ...798...10R,2018ApJ...852...53R}.
}

\section{The Luminosity power-law behavior in BdHN  measured in the rest-frame}\label{sec:8}

In the following, we are going to fit simultaneously the light-curves of all the $21$ BdHNe with GeV emission {expressed in the rest-frame of the source.} We assume a same power-law decay index but different amplitudes, this assumption is consistent with our model, moreover, it brings the benefit for those GRBs with limited data that cannot be fitted solely.

We limit our analysis of the light-curves at time later than the BdHN prompt emission, when the GeV luminosity is already in the asymptotic power-law regime. A power-law
\begin{equation}
	L_n(t) = A_n t^{\alpha},
\end{equation}
describing the rest-frame $0.1$--$100$~GeV isotropic luminosity light-curve of $n$th BdHN is assumed. In the simultaneous fitting, we perform the Levenberg-Marquardt method to perform the minimization \citep{Levenberg-Marquardt}. The basic idea of fitting  is to minimize the $\chi^2$; when fitting one curve to one equation, the $\chi^2$ is minimized. To fit $N$ curves to $N$ equations simultaneously, the sum of the $\chi^2$ values should to be minimized. The related equations are:
\begin{align}
\chi^2 &= \sum_{n=1}^{N}~\chi^2 _n,\label{eq1} \\
\chi^2 _n &= \sum_{i=1}^{M} \frac{1}{\sigma_{ni}^2}(L_{ni}-L_n(t_{ni},A_n, \alpha))^2,\label{eq2}
\end{align}
where $n$ represents each BdHN, $i$ represents each data point in a given BdHN, $A_n$ is the amplitude of a power-law function for the $n$th BdHN, $\alpha$ is the common power-law index shared for all the BdHNe. Thus, for the $n_{\rm th}$ BdHN, at time $t_{ni}$, the observed luminosity is $L_{ni}$, and the predicted luminosity is $L_n(t_{ni},A_n, \alpha)$. The value of $\chi^2$ represents the difference between the best power-laws fitting and all the observed data, it is a summation of individual $\chi^2 _n$, which represents the difference between the power-law fitting and the observed value of each GRB.

\begin{figure*}
\centering
\includegraphics[width=1\hsize,clip]{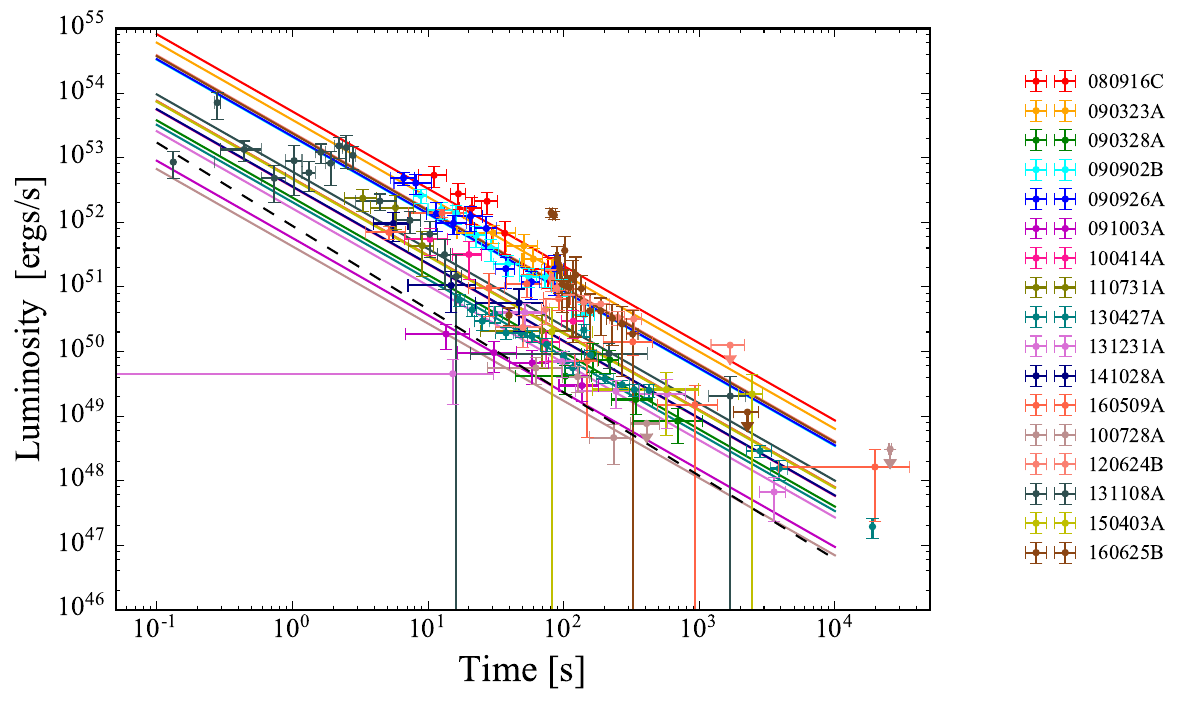}
\caption{The rest-frame $0.1$--$100$~GeV isotropic luminosity light-curves of $17$ selected BdHNe with LAT emission. The solid red line marks the common power-law behavior of the GeV emission for BdHNe with slope $\alpha=1.20\pm0.04$; the shaded gray area encloses all the luminosity light-curves of the selected BdHNe. The dashed black line  marks the common power-law behavior of the GeV emission in S-GRBs with slope $\gamma=1.29 \pm0.06$.}
\label{fig:02}
\end{figure*}

Out of $21$ BdHNe presented in Table~\ref{tab:cb} we perform the fitting for only $17$ GRBs which have more than two data points in their luminosity light-curves. Therefore, for the fitting of BdHNe, there are $17$ bursts and each one has its power-law function. Consequently, there are in total $17$ parameters, including $17$ amplitudes, and $1$ power-law index. The fitting gives the power-law index $\alpha = 1.2 \pm 0.04$, i.e.:
\begin{equation}
L_{n}=A_{n}~t^{~-1.2 \pm 0.04}, \label{eq3}
\end{equation}
which is plotted in Fig.~\ref{fig:02} and the amplitudes of each GRB, $A_{n}$, with the uncertainty are shown in Table~\ref{tab:23fit}. {This inferred  power law index is similar one obtained from fitting of GeV flux, $f_{\nu}(t)$, see e.g. \citep{2009MNRAS.400L..75K} and \citep{2017ApJ...837...13P} , in which power-law index is $\alpha=1.2 \pm 0.2$ and $\alpha=1.2 \pm 0.4$, respectively. Since luminosity is proportional to flux, i.e.~$L= 4 \pi d^2_L (1+z)^{\alpha -2} f_{\mu}$ which $d_L$ is luminosity distance, this similarity of power-law index is not surprising. The advantage of using luminosity expressed in the rest-frame of the source, instead of flux in arrival time, is that one can determine the intrinsic energy loss of the system which produces the GeV radiation, regardless of difference in the redshift of their sources.}

\begin{table*}
\centering
\begin{tabular}{ccccc}
\hline\hline
BdHNe & $A_{n}$ (Amplitude) & uncertainty of $A_{n}$  & $L_{10s}$ & uncertainty of $L_{10s}$\\
\hline
\hline
080916C & $5.201\times10^{53}$ & $^{+1.605}  _{-1.307} \times10^{53}$ & $3.341\times10^{52}$ & $^{+1.963}  _{-1.800} \times10^{52}$ \\
090323A & $3.847\times10^{53}$ & $^{+1.436}  _{-1.169} \times10^{53}$ & $2.472\times10^{52}$ & $^{+1.542}  _{-1.397} \times10^{52}$ \\
090328A & $2.408\times10^{52}$ & $^{+1.087}  _{-0.773} \times10^{52}$ & $1.547\times10^{51}$ & $^{+1.042}  _{-0.889} \times10^{51}$  \\
090902B & $2.091\times10^{53}$ & $^{+5.845}  _{-4.599} \times10^{52}$ & $1.343\times10^{52}$& $^{+7.696}  _{-7.055} \times10^{51}$\\
090926A & $2.141\times10^{53}$ & $^{+5.887}  _{-4.838} \times10^{52}$ & $1.376\times10^{52}$ & $^{+7.850}  _{-7.259} \times10^{51}$ \\
091003A & $5.715\times10^{51}$ & $^{+1.735}  _{-1.520} \times10^{51}$ & $3.671\times10^{50}$ & $^{+2.147}  _{-2.004} \times10^{50}$ \\
100414A & $3.529\times10^{52}$ & $^{+1.399}  _{-1.142} \times10^{52}$ & $2.267\times10^{51}$ & $^{+1.446}  _{-1.306} \times10^{51}$\\
100728A & $4.241\times10^{51}$ & $^{+1.978}  _{-1.512} \times10^{51}$ & $2.725\times10^{50}$ & $^{+1.863}  _{-1.622} \times10^{50}$ \\
110731A & $4.807\times10^{52}$ & $^{+1.707}  _{-1.442} \times10^{52}$ & $3.088\times10^{51}$& $^{+1.894}  _{-1.739} \times10^{51}$ \\
120624B & $2.459\times10^{53}$ & $^{+8.261}  _{-6.167} \times10^{52}$ & $1.580\times10^{52}$ & $^{+9.518}  _{-8.513} \times10^{51}$\\
130427A & $2.053\times10^{52}$ & $^{+5.124}  _{-4.091} \times10^{51}$ & $1.318\times10^{51}$ & $^{+7.370}  _{-6.815} \times10^{50}$\\
131231A & $1.637\times10^{52}$ & $^{+7.818}  _{-5.446} \times10^{51}$ & $1.052\times10^{51}$& $^{+7.273}  _{-6.116} \times10^{50}$ \\
141028A & $3.590\times10^{52}$ & $^{+1.225}  _{-1.109} \times10^{52}$ & $2.306\times10^{51}$& $^{+1.396}  _{-1.310} \times10^{51}$ \\
131108A & $6.077\times10^{52}$ & $^{+9.089}  _{-8.894} \times10^{51}$ & $3.904\times10^{51}$  & $^{+2.037}  _{-1.947} \times10^{51}$\\
150403A & $4.671\times10^{52}$ & $^{+2.034}  _{-1.595} \times10^{52}$ & $3.001\times10^{51}$ & $^{+1.989}  _{-1.760} \times10^{51}$\\
160509A & $4.812\times10^{52}$ & $^{+1.733}  _{-1.313} \times10^{52}$ & $3.091\times10^{51}$& $^{+1.905}  _{-1.698} \times10^{51}$ \\
160625B & $2.378\times10^{53}$ & $^{+8.093}  _{-5.854} \times10^{52}$ & $1.528\times10^{52}$ & $^{+9.241}  _{-8.199} \times10^{51}$\\
\hline
\hline
\end{tabular}
\caption{\textit{{Fitting parameters of the relation between $0.1$--$100$~GeV luminosity vs. time when measured in the rest frame of 17 BdHNe with GeV emission}}: amplitude of the BdHNe $0.1$--$100$~GeV luminosity, $A_n$, and its uncertainty, the inferred $0.1$--$100$~GeV luminosity at $10$~s from the fitting and its uncertainty. The common power-law index is $\alpha= 1.20\pm 0.04$. Out of $21$ BdHNe emitting GeV emission we performed the fitting for $17$ GRBs which have more than two data points in their luminosity light-curves. {GRBs 091208B, 130518A, 150314A, 150514A have only two data points in their GeV luminosity light curves}.} 
\label{tab:23fit}
\end{table*}

\begin{figure*}
\centering
\includegraphics[width=0.49\hsize,clip]{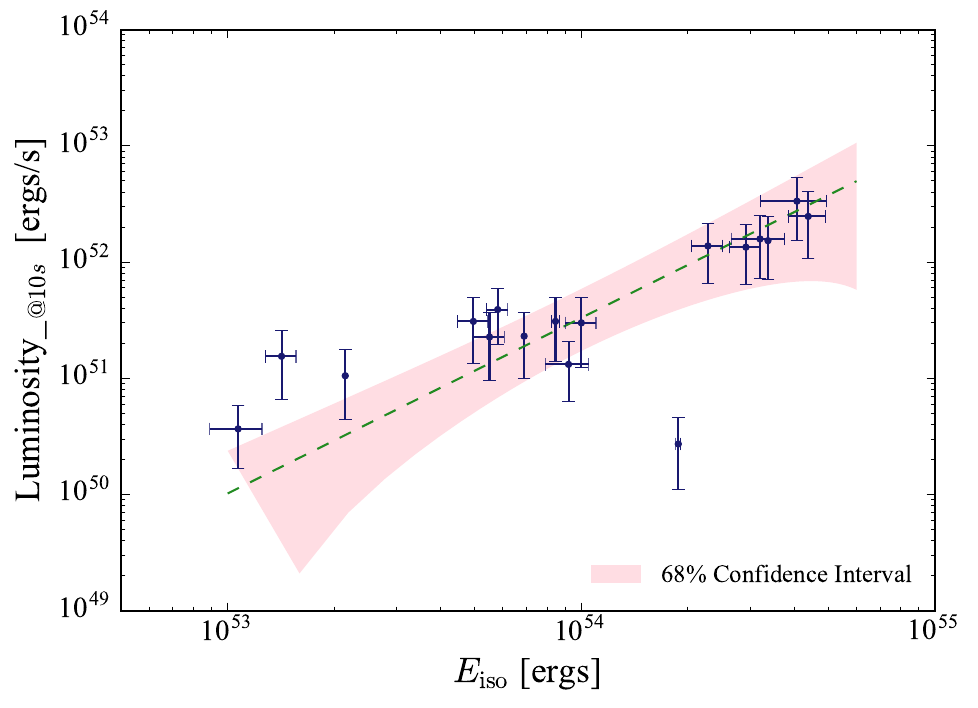}
\includegraphics[width=0.49\hsize,clip]{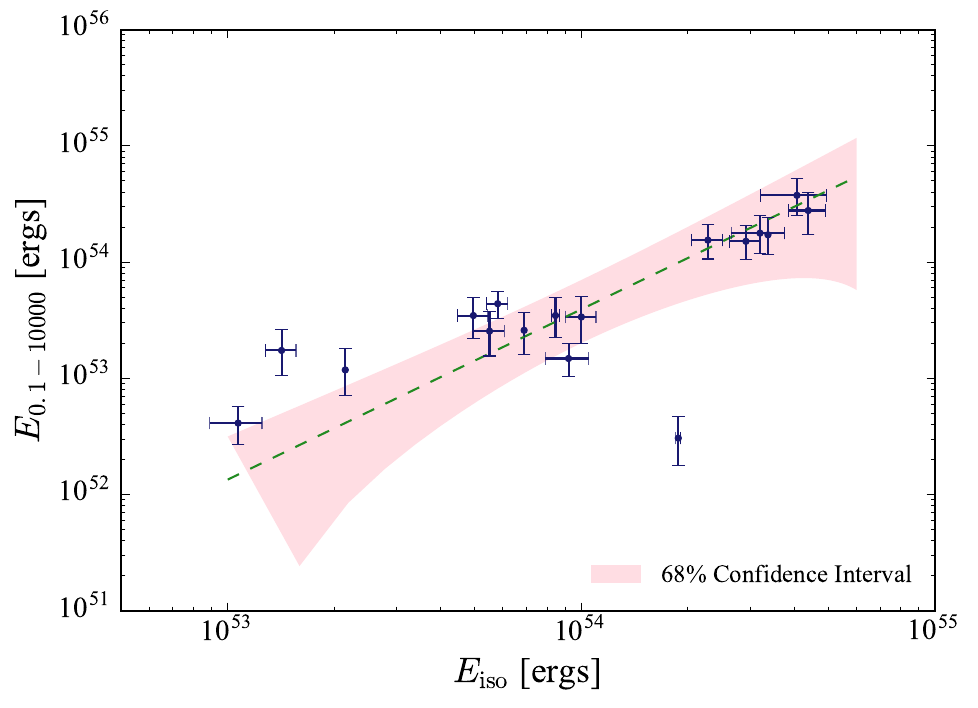}
\caption{{Left}: the \textit{Fermi}-LAT luminosity at $10$~s in the energy range $0.1$--$100$~GeV versus the isotropic gamma-ray energy from $1$~keV to $10$~MeV. The BdHNe are listed in Table~\ref{tab:23fit}. {Right}: the \textit{Fermi}-LAT energy from $0.1$ to $10^4$~s versus isotropic gamma-ray energy from $1$ keV to $10$~MeV. {See the corresponding values in Table~\ref{tab:14fit}.}}
\label{K10Eiso}
\label{EEiso}
\end{figure*}

After having the best power-law parameters of the light curve for each BdHNe, we check the correlation between the GeV luminosity at $10$~s from Eq.~(\ref{eq3}) using the fitted parameters and the isotropic energy $E_{\rm iso}$. The power-law fitting gives (see Fig.~\ref{K10Eiso}):
\begin{equation}
L_{\mathrm{10s}}=(3.701~\pm 2.21) \times 10^{51}~E_{\mathrm{iso}}^{~1.428 \pm 0.425}, \label{eq4}
\end{equation}
and the fitting parameters for each GRB including their uncertainties are shown in Table~\ref{tab:23fit}. Furthermore, we estimate the energy released in the GeV band by each GRB in the $0.1$--$10^{4}$~s time interval, i.e.:
\begin{equation}
E_{\mathrm{0.1-10^{4}s}}=A_{\mathrm{GRB}}~\int_{0.1}^{10000}t^{-1.19}~dt\label{eq6}
\end{equation}
and the derived $E_{\mathrm{0.1-10^{4}s}}$ are shown in Table~\ref{tab:14fit}. The parameters $E_{\mathrm{0.1-10^{4}s}}$ and $E_{\rm iso}$ (isotropic energy of the prompt emission in $\gamma$ band) are also correlated by a power-law relation (see Fig.~\ref{EEiso}):
\begin{equation}
E_{\mathrm{0.1-10^{4}s}}=(4.125~\pm 2.59) \times 10^{53}~E_{\mathrm{iso}}^{~1.424 \pm 0.447}.\label{eq7}
\end{equation}

\begin{table*}
\centering
\begin{tabular}{cccc}
\hline\hline
BdHNe & $E_{0.1-10^{4}s}$ & positive uncertainty of $E_{0.1-10^{4}s}$ & negative uncertainty of $E_{0.1-10^{4}s}$\\
\hline
\hline
080916C & $3.752\times10^{54}$ & $1.460\times10^{54}$ & $1.242\times10^{54}$ \\
090323A & $2.775\times10^{54}$ & $1.227\times10^{54}$ &   $1.034\times10^{54}$\\ 
090328A & $1.737\times10^{53}$ & $8.860\times10^{52}$ &   $6.723\times10^{52}$\\
090902B & $1.508\times10^{54}$ & $5.529\times10^{53}$ &   $4.646\times10^{53}$\\
090926A & $1.545\times10^{54}$ & $5.608\times10^{53}$ &    $4.824\times10^{53}$\\
091003A & $4.122\times10^{52}$ & $1.588\times10^{52}$ &    $1.411\times10^{52}$\\
100414A & $2.546\times10^{53}$ & $1.176\times10^{53}$ &    $9.902\times10^{52}$\\
100728A & $3.060\times10^{52}$ & $1.600\times10^{52 }$ &   $1.275\times10^{52}$\\
110731A & $3.467\times10^{53}$ & $1.481\times10^{53}$ &   $1.281\times10^{53}$\\
120624B & $1.774\times10^{54}$ & $7.294\times10^{53}$ &   $5.867\times10^{53}$\\
130427A & $1.481\times10^{53}$ & $5.098\times10^{52}$&   $4.348\times10^{52}$\\
131108A & $4.383\times10^{53}$ & $1.228\times10^{53}$ &   $1.142\times10^{53}$\\
131231A & $1.181\times10^{53}$ & $6.297\times10^{52}$ &   $4.682\times10^{52}$\\
141028A & $2.589\times10^{53}$ & $1.076\times10^{53}$ &   $9.757\times10^{52}$\\
150403A & $3.369\times10^{53}$ & $1.671\times10^{53}$ &   $1.361\times10^{53}$\\
160509A & $3.471\times10^{53}$ & $1.497\times10^{53}$ &   $1.207\times10^{53}$\\
160625B & $1.716\times10^{54}$ & $7.116\times10^{53}$ &   $5.614\times10^{53}$\\
\hline
\hline
\end{tabular}
\caption{Results of $E_{0.1-10^{4}s}$ and related error of 17 BdHNe. $E_{0.1-10^{4}s}$ is total GeV energy (in erg) emitted from $0.1$ to $10^4$~s. {GRBs 091208B, 130518A, 150314A, 150514A are excluded since they have only two data points in their GeV luminosity light curves.}} 
\label{tab:14fit}
\end{table*}

This positive correlations indicates that the BdHNe with higher isotropic energy are also more luminous and more energetic in the GeV emission. 

\section{Spin down of the BH in BdHNe}\label{sec:9}

{We can turn now from the luminosity expressed in the rest-frame of the sources, see Eq.~(\ref{eq3}) and from the initial values of the spin and mass of the BH expressed in section~\ref{sec:5} to derive the slowing down of the BH due to the energy loss of in the GeV emission. The time derivative of Eq.~(\ref{Eextr}) gives the luminosity}
\begin{equation}
\label{sdown1}
L=-\frac{dE_{extr}}{dt}=-\frac{dM}{dt},
\end{equation}
{Since $M_{irr}$ is constant for each BH during the energy emission process, and using our relation for luminosity $L=A t^{-1.2}$, we obtain the relation of the loss of mass-energy of the BH by integrating Eq.~(\ref{sdown1}):}
\begin{equation}
\label{sdown2}
M= M_0 + 5 A t^{-0.2}-5A t_0^{-0.2},
\end{equation}
{which $M_0$ is the initial mass of newly born BH. From the mass-energy formula of the BH we have}
\begin{equation}
\label{sdown3}
J= 2 M_{irr} \sqrt{M^2-M^2_{irr}},
\end{equation}
{therefore}
\begin{equation}
\label{sdown4}
a=\frac{J}{M}= 2 M_{irr} \sqrt{1-\frac{M^2_{irr}}{(M_0 + 5 A t^{-0.2}-5A t_0^{-0.2})^2}}.
\end{equation}

\begin{figure*}
\centering
\includegraphics[width=0.85\hsize,clip]{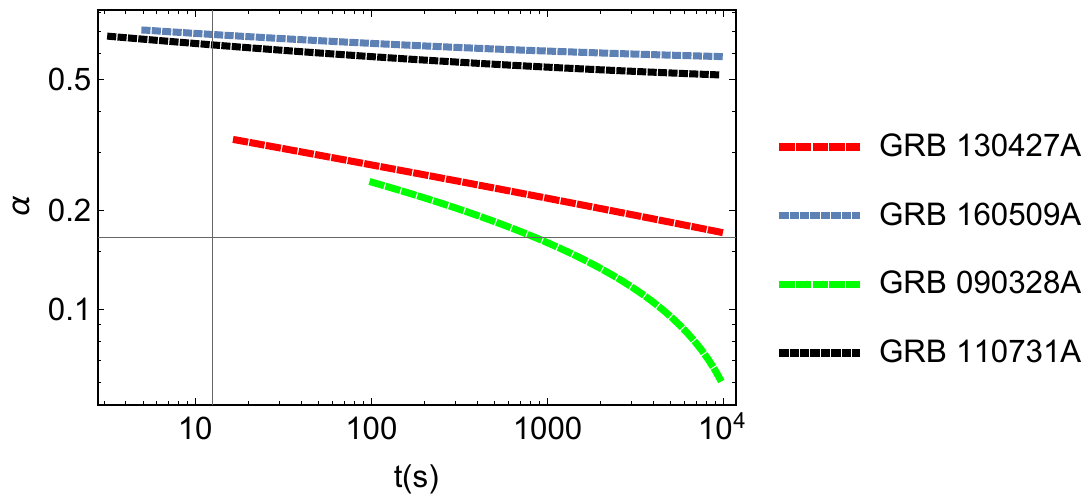}
\caption{{The decrease of the BH spin, as a function of rest-frame time. The initial values of spin and mass within the TM1 equation of state for GRB 090328A are $\alpha=0.2434$ and $M(\alpha)=2.2526 M_{\odot}$, for 110731A: $\alpha=0.68$ and $M(\alpha)=2.57 M_{\odot}$
, for 130427A: $\alpha=0.327$ and $M(\alpha)=2.2893 M_{\odot}$
and for 160509A: $\alpha=0.707$ and $M(\alpha)=2.636 M_{\odot}$. The initial values of spin and mass within the NL3 model are also shown in Table.~\ref{tab:dd}. This behavior of spin parameter indicates that rotational energy of the BH is decreasing due to the radiation losses in GeV.}}
\label{fig:sdown}
\end{figure*}

{As indicative examples we show in Fig.~\ref{fig:sdown} the decrease of the BH spin, $\alpha=a/M=J/M^2$, as a function of time in GRBs 090328A, 110731A, 130427A and 160509A.}

\section{Conclusions}\label{sec:12}

{On the ground of recent progress in the classification of  GRBs in eight different families, reviewed in section~\ref{sec:2}, we have examined the role of the GeV emission in long GRBs. The goal is to further extend the observables in this GRB classification and follow a new approach to explain  the GeV emission by assuming that it  originates from the rotational energy of a Kerr-Newman BH.}

{From the theoretical point of view we have relied on our previous works, of the mass-energy formula of the BH on the theory of hypercritical accretion and on the 3D simulation performed with an SPH code to construct the morphology of the GRB.}

{We have given attention to the subclass of BdHNe, long GRBs originating in a binary system composed of a CO$_{core}$, undergoing a SN explosion, and a companion NS. It had already been established  that when the binary period is longer than $5$ minutes the hypercritical accretion of the SN ejecta is not sufficient to overcome its critical mass, and no BH is formed. An XRF with $E_{iso}<10^{52}$~erg occurs, consistently we have verified in section \ref{sec:3} that in $48$ XRFs no GeV emission occurs: No BH No GeV emission.}

{For a shorter binary period, the hypercritical accretion is sufficient to lead the NS to overcome its critical mass and collapse to a BH. This originates a BdHNe with $E_{iso}>10^{52}$~erg. We have addressed both BdHNe with BH, either newly-formed or preexisting.}

{In order to ascertain the presence of the GeV emission, out of $329$ BdHNe, in section~\ref{sec:4} we have focused on $48$ within the boresight angle of the \textit{Fermi}-LAT; we have found $21$ with and $27$ without GeV emission. For each of the $21$ sources we have given the basic parameters in Table~\ref{tab:cb}.}

{In section~\ref{sec:5} we have implemented our basic assumption that the GeV emission originates from the rotational energy of a Kerr-Newman BH. Consequently, we have determined for each source the initial spin and mass of the BH to fulfill the energetic requirement of the GeV radiation.}

{\textit{The first new conclusion} is that there are two types of BdHNe. $16$ BdHNe of type one (BdHNe I) have a BH originating from the gravitational collapse of a NS, see Table~\ref{tab:dd}, and $5$ BdHNe of type two (BdHNe II) originate from an already existing BH and are the most energetic BdHNe, see Table~\ref{tab:2b}. For all of them the BH rotational energy is sufficient to explain the energetic of all GeV emissions.}

{We have then addressed the issue of the $27$ BdHNe without the GeV emission and using specific examples in our previous works we have assumed  that these sources have a viewing angle laying in the orbital plane of the binary progenitor system. When the \textit{Swift} data are available  hard and soft X-ray flare as well as extended thermal emissions are observed.}

{\textit{The second new conclusion} is that the GeV emission occurs in a cone with an half-opening angle of approximately $60^\circ$ from the normal to the plane of the binary progenitor. This evidence  a new asymmetric morphology of the BdHNe reminiscent of the one observed from AGN and well illustrated by the reconstruction of the 3D SPH code that follows the SN ejecta expansion accreting onto the NS companion shown in Fig.~\ref{fig:cc}. We have also verified that the condition of transparency of the GeV emission requires a Lorentz factor $\Gamma \approx  1500$ for the source of the GeV emission.}

{\textit{The third new conclusion:} in section~\ref{sec:8} we have implemented  the description of the luminosity of the GeV emission as a function of time in the rest-frame of the source and found a universal power-law dependence with index of $\alpha = 1.2 \pm 0.04$. This has allowed to compute for each source the slowing down rate of the BH spin due to the GeV radiation.
In this and previous papers we have shown that the Kerr-Newman solution can fulfill the energy requirement of the GeV emission in both long and short GRBs and consequently to determine the BH mass and spin. We are currently analyzing the electrodynamical properties of the Kerr-Newman BH for studying the origin of the high-energy process in particle and in fields common in GRBs and AGN.}

{We have seen evidence of an ultra-relativistic emission originating from a BH, for long GRBs, in this article and, for short GRBs, in previous articles. However, this does not occur in all GRBs as traditionally assumed. It occurs only in a subclass of BdHNe and in a subclass of short GRBs. In both cases this ultra-relativistic emission, occurs in the GeV energy range, and has an extended luminosity following precise power-laws when expressed in the source rest-frame. This emission is additive to the other components of the GRB which include, in the gamma-rays, an additional ultra-relativistic component in the prompt radiation phase lasting $\lesssim 20$~s and, in the X-rays, an additional mildly-relativistic phase with Lorentz factor $\Gamma \sim 2$ extending to the plateau and the afterglow phase.}

\acknowledgments
We acknowledge the protracted discussion with Roy Kerr. We also acknowledge the continuous support of the MAECI and the Italian Space Agency (ASI). This work made use of data from Fermi space observatory. Y.~A. is supported by the Erasmus Mundus Joint Doctorate Program Grant N.~2014-0707 from EACEA of the European Commission. Y.~A. acknowledges partial support by the Targeted Financing Program BR05336383 of Aerospace Committee of the Ministry of Defense and Aerospace Industry of the Republic of Kazakhstan. G.~J.~M is supported by the U.S. Department of Energy under Nuclear Theory Grant DE-FG02-95-ER40934. M.~M. acknowledges partial support provided by a targeted funding for scientific and technical program of the Ministry of Education and Science of Kazakhstan.

\appendix

\section{Optically thin condition for the highest energy photon in $\gamma+\gamma \rightarrow e^{+}e^{-}$ collision}\label{apd:transparency}

Assuming the photons having energy more than the most energetic photon  observed are annihilated by the collisions with the lower energy photons and produce electron-positron pairs as  $\gamma+\gamma \rightarrow e^{+}e^{-}$. This assumption is equivalent to the expression of the optical depth 
\begin{equation}
\tau(\epsilon_{max}) = \sigma  n(\epsilon>\epsilon_{max,th}) \Delta r = 1
\end{equation} 
where $\epsilon_{max}$ is the highest energy of photon observed,  $\epsilon_{max,th}$ is the corresponding threshold of energy for having the pair creation, $n(\epsilon>\epsilon_{max,th})$ is the number density of photon more the threshold energy at radius $r$, $\Delta r$ is the  thickness of the plasma.

We are going to find all the terms in the above equation. The average cross section of gamma-ray annihilation is $\sigma = \frac{11}{180} \sigma_T$ \citep{1987MNRAS.227..403S}. The threshold energy for photon of $\epsilon_{max}$ in the observer's frame gives
\begin{equation}
\epsilon_{max,th} = \frac{(\Gamma m_e c^2)^2}{(1+z)^2 \epsilon_{max}},
\end{equation} 
where $\Gamma$ is the bulk Lorentz factor of the plasma outflow, $z$ is the redshift. The density $n$ can be obtained from the observation. The typical $\epsilon_{max,th}$ is more than $10~$ MeV for the Fermi-LAT photons from a GRB, therefore, the annihilation is mainly caused by the ultra-high energy photons themselves, not from the low energy photons possibly from other regions. For example, at time  $t$ in the observer's frame, the fitted count flux density follows a power-law 
\begin{equation}
f = f_0 \epsilon^{-\beta}
\end{equation} 
in units of count~cm$^{-2}$~s$^{-1}$~erg$^{-1}$, the index $\beta \geq 2$ for most of the cases. The density at the observer is 
\begin{eqnarray}
n_{obs}(\epsilon>\epsilon_{max,th}) &=& c^{-1} \int_{\epsilon_{max,th}}^{\infty} f_0 \epsilon^{-\beta} d\epsilon  \\ &=& \frac{f_0}{c (\beta-1)} (1+z)^{2(\beta-1)}(\Gamma m_e c^2)^{-2(\beta-1)} \epsilon_{max}^{\beta-1}
\end{eqnarray} 

Assuming $\Gamma$ is constant, the distance of the plasma at time $t$ is
\begin{equation}
r \simeq 2 \Gamma^2 c t,
\end{equation} 
and the density at radius $r$ is
\begin{equation}
n = \frac{1}{(1+z)^2} \frac{D_L^2}{r^2}  n_{obs} \\ = \frac{f_0}{4 c^3 (\beta-1)} (m_e c^2)^{-2(\beta-1)} \epsilon_{max}^{\beta-1} D_L^2 (1+z)^{2(\beta-2)} \Gamma^{-2(\beta+1)} t^{-2}.
\end{equation} 

By assuming $\Delta r \simeq r$, we have the final optical depth 
\begin{eqnarray}
\tau &=& \sigma n \Delta r \\ &=& \frac{ 11 f_0 \sigma_T}{360 c^2 (\beta-1)} (m_e c^2)^{-2(\beta-1)} \epsilon_{max}^{\beta-1} D_L^2 (1+z)^{2(\beta-2)} \Gamma^{-2 \beta} t^{-1}.
\end{eqnarray} 

The Lorentz factor can be obtained by equating the optical depth to 1, then we have
\begin{equation}
\Gamma = (\frac{ 11 f_0 \sigma_T}{360 c^2 (\beta-1)})^{1/2\beta} (m_e c^2)^{-2(\beta-1)/2\beta} \epsilon_{max}^{(\beta-1)/2\beta} D_L^{1/\beta} (1+z)^{(\beta-2)/\beta}  t^{-1/2\beta}.
\end{equation} 

In the case of $\beta = 2$,
\begin{eqnarray}
\Gamma_{\beta=2} &=& (\frac{ 11 f_0 \sigma_T}{360 c^2})^{1/4} (m_e c^2)^{-1/2} \epsilon_{max}^{1/4} D_L^{1/2} t^{-1/4} \\ &=& 1525.37 ~(\frac{f_0}{10^{-4}})^{1/4} (\frac{D_L}{10^{27} cm})^{1/2} (\frac{\epsilon_{max}}{10 GeV})^{1/4}(\frac{t}{10s})^{-1/4}.
\end{eqnarray} 

For $f_0=10^{-4}$ and $D_L = 10^{27}$~cm, corresponding to luminosity $\sim 10^{52}$~erg~s$^{-1}$, and at $t=10$~s, $\epsilon_{max} = 10$~GeV, the inferred $\Gamma \sim 1500$, and distance $r \sim 10^{18}$~cm.

The above consideration is similar to \citet{2001ApJ...555..540L}, but it is derived to be applicable for the GeV region, and the parameters fitted from the observed spectrum.

\end{document}